\documentclass{aa}
\usepackage{txfonts,graphicx,longtable}

\newcommand{\mtwosini}{$m_2\,\sin i$}
\newcommand{\mjup}{M$_{\mbox{\tiny Jup}}$}

\newcommand{\masa}{mas$\cdot$yr$^{-1}$}

\newcommand{\msun}{M$_{\odot}$}
\newcommand{\chisq}{$\chi^2$}

\newcommand{\epseri}{$\epsilon$~Eri}

\begin{document}

\title{Mass constraints on substellar companion candidates from the re-reduced 
{\it Hipparcos} intermediate astrometric data:
Nine confirmed planets and two confirmed brown dwarfs
\thanks{Based on observations collected by the {\it Hipparcos} satellite}}

\author{Sabine Reffert \and Andreas Quirrenbach}

\institute{ZAH - Landessternwarte, K\"onigstuhl~12, 69117~Heidelberg, Germany}

\date{Received <date> / Accepted <date>}

\abstract{The recently completed re-reduction of the \begin{it}Hipparcos\end{it} data by van Leeuwen (2007a)
makes it possible to search for the astrometric signatures of planets and brown
dwarfs known from radial velocity surveys in the improved 
\begin{it}Hipparcos\end{it} intermediate astrometric data.}
{Our aim is to put more significant constraints on the orbital parameters which cannot be 
derived from
radial velocities alone, i.e.\ the inclination and the longitude of the ascending 
node, than was possible before. The determination
of the inclination in particular allows to calculate an unambiguous companion mass,
rather than the lower mass limit which can be obtained from radial velocity measurements.}
{We fitted the astrometric orbits of 310 substellar companions around 258 stars,
which were all discovered via the radial velocity method, to the \begin{it}Hipparcos\end{it} 
intermediate astrometric data provided by van Leeuwen.}
{Even though the astrometric signatures of the companions cannot be detected in
most cases, the \begin{it}Hipparcos\end{it} data still provide lower limits on the inclination 
for all but 67 of the investigated companions, which translates into upper limits 
on the masses of the unseen companions.
For nine companions the derived upper mass limit lies in the planetary and for 
75 companions in the brown dwarf mass regime, proving the substellar nature of those 
objects. Two of those objects have minimum masses also in the brown dwarf regime and
are thus proven to be brown dwarfs.
The confirmed planets are the ones around Pollux ($\beta$~Gem~b), \epseri~b,
$\epsilon$~Ret~b, $\mu$~Ara~b, $\upsilon$~And~c and d, 
47 UMa~b, HD~10647~b and HD~147513~b. 
The confirmed brown dwarfs are HD~137510~b and HD~168443~c.
In 20 cases, the astrometric signature of the substellar companion was
detected in the \begin{it}Hipparcos\end{it} data, resulting in reasonable constraints on inclination and
ascending node. Of these 20 companions, three are confirmed as planets or lightweight brown
dwarfs (HD~87833~b, $\iota$~Dra~b, and $\gamma$~Cep~b), two as brown dwarfs
(HD~106252~b and HD~168443~b), and four are low-mass stars (BD\,--04~782~b, HD~112758~b,
$\rho$~CrB~b, and HD169822~b). Of the others, many are either 
brown dwarfs or very low mass stars.
For \epseri, we derive a solution which is very similar to the one obtained using \begin{it}Hubble\end{it}
Space Telescope data.
}
{}


\titlerunning{Mass constraints on known substellar companions from the re-reduced {\it Hipparcos} data}
\authorrunning{Reffert \& Quirrenbach}

\maketitle

\section{Introduction}
\label{intro}

Most of the around 400 planets and planet candidates known so 
far\footnote{see http://www.exoplanets.org/}
have been detected with the radial velocity method. As is very well known, 
only five of the seven orbital parameters which characterize a binary orbit
can be derived from radial velocity observations (period, periastron time,
eccentricity, longitude of periastron and either the mass function, the radial 
velocity semi-amplitude, the semi-major axis of the companion orbit or the semi-major
axis of the primary orbit times the sine of the inclination). The remaining
two orbital parameters, inclination and longitude of the ascending node\footnote{In 
the following, we denote this orbital element with `ascending node', short for 
`longitude of the ascending node'}, are related to the
orientation of the orbit in space and cannot be derived from radial velocities 
alone. Unfortunately, this means that the true companion mass is usually not
known for planets detected via radial velocities; the only quantity which can
be derived is a lower limit for the companion mass, which is the true companion
mass times the sine of the inclination. Strictly speaking, all substellar 
companions detected via the radial velocity technique would thus have to be
called planet or brown dwarf {\it candidates} since their true mass is not known. 

In order to establish substellar companion masses, a complementary technique 
is required which is sensitive to the inclination. For companions
with inclinations close to 90\degr, transit photometry can provide the
inclination and thus, among other information, an accurate companion mass. For most
systems however this is not the case, and astrometry is the method of choice.
Astrometry provides both missing parameters, the inclination and the ascending node,
so that the orbits are fully characterized. In principle, all orbital
parameters can be derived from astrometric measurements, but since with
current techniques the radial velocity parameters are usually more precise,
another option is to derive only the two missing orbital parameters from the
astrometry. 

The astrometric signatures of planetary companions are rather small
compared to the astrometric accuracies which are currently achievable.
The astrometric signature $\alpha_{\mbox{\scriptsize max}}$ can be calculated 
if the radial velocity semi-amplitude $K_1$, the period $P$, the
eccentricity $e$, the inclination $i$ and the parallax $\varpi$ are known
(see e.g.\ Heintz \cite{heintz78} or any other book on double stars):
\begin{equation}
\label{eqastsig}
\alpha_{\mbox{\scriptsize max}} = \frac{K_1\cdot P\cdot \sqrt{1-e^2}\cdot\varpi}{1~\mbox{AU}\cdot2\pi\cdot \sin i}
\end{equation}
$\alpha_{\mbox{\scriptsize max}}$ corresponds to the semi-major axis of the true orbit of the 
photocenter around the center of mass. This is however in general not the astrometric
signature which is observable from Earth in the case of eccentric orbits,
due to projection effects. In the worst case, the semi-major axis of the
projected orbit corresponds to the semi-minor axis of the true orbit only,
while the semi-minor axis of the projected orbit could be identical to zero, so
that no astrometric signal can be observed along that direction.
The astrometric signature $\alpha$ which is actually observable from Earth can
be calculated with the formulae for ellipse projection given in Appendix~\ref{apparent}.

If the inclination is unknown, as is the case when the orbit is derived
from radial velocities alone, 
Eq.~\ref{eqastsig} provides a lower limit
for the astrometric signature by setting $\sin i = 1$ and multiplying with
$1-e^2$ (converting the semi-major into the semi-minor axis):
\begin{equation}
\alpha_{\mbox{\scriptsize min}} = \frac{K_1\cdot P\cdot (1-e^2)^{3/2}\cdot\varpi}{1~\mbox{AU}\cdot2\pi}
\end{equation}
We note that a
more stringent lower limit on the observable astrometric signature could be obtained if the 
longitude of the periastron is taken into account; the above lower limit for the astrometric 
signature corresponds to the case where the longitude of the periastron is $\pm 90$\degr.

Instead of the semi-amplitude $K_1$, one could also use the
mass function or the semi-major axis of the primary orbit, multiplied by
$\sin i$, as an alternative orbital element to express the unique relation 
between astrometric signature and various orbital elements.

Equation~\ref{eqastsig} shows that we can always derive an upper mass limit
for the companion based on astrometry if the companion mass times the sine of
the inclination, $m_2\cdot\sin i$, is known from radial velocities. The astrometry 
provides an upper limit on the astrometric signature of the companion, 
which translates into a lower limit on the inclination (Eq.~\ref{eqastsig}).
In combination with the minimum companion mass
$m_2\cdot\sin i$, the lower limit on the inclination translates into an upper
limit of the companion mass $m_2$.

For example, the astrometric signature of a
companion with a mass of 1~\mjup\ and a period of five years
orbiting around a solar-mass star located at a distance of 10~pc 
amounts to 0.28~mas, not taking projection effects into account.
Although this value is rather small, several astrometric 
detections of substellar
companions have been reported in the past, mostly for systems with
rather long periods and relatively massive companions around nearby stars,
which all help to increase the astrometric signal. 

Based on the {\it Hipparcos} data, \cite{perryman96} derived upper mass limits
of 22~\mjup\ for the substellar companion to 47~UMa and of 65~\mjup\ for
the companion to 70~Vir, with 90\% confidence.
This confirmed for the first time the substellar nature
of these newly detected companions. For 51~Peg with its period of only 
a few days no useful upper mass limit could be established based on 
{\it Hipparcos} data. 

Mazeh et al.\ (\cite{mazeh99}) and \cite{zucker00} followed that same approach 
to derive masses or upper mass limits for the companions to $\upsilon$~And and
HD~10697, respectively. For the outermost companion in the $\upsilon$~And
system, Mazeh et al.\ (\cite{mazeh99}) derived a mass of 10.1$^{+4.7}_{-4.6}$~\mjup\ 
at a confidence level of 68.3\% and a mass of 10.1$^{+9.5}_{-6.0}$~\mjup\ at a
confidence level of 95.4\%. For the companion to HD~10697, \cite{zucker00}
obtained a mass of 38$\pm$13~\mjup, which implies that the companion is
actually a brown dwarf and not a planet. These studies were extended to
all the 47 planetary and 14 brown dwarf companion candidates known at the time in
\cite{zucker01}. For 14~planetary companions, the derived upper mass limits
imply that the companions are of substellar nature, but for the others
no useful upper mass limits could be derived. Similarly, even for the brown
dwarfs the {\it Hipparcos} astrometry is in general not precise enough to derive
tight upper limits or to establish the astrometric orbit. However, for six
of the 14 brown dwarf candidates it turned out that the companion was stellar,
and the astrometric orbit could be derived. This confirms the results of
Halbwachs et al.\ (\cite{halbwachs00}), who examined the {\it Hipparcos} astrometry for eleven 
stars harboring brown dwarf candidates. Seven of those brown dwarf secondaries
turned out to be of stellar mass, while only one of the studied companions 
is, with low confidence, a brown dwarf. For the other three candidates, no
useful constraints could be derived. In a previous paper (Reffert \& Quirrenbach
\cite{reffert06b}),
we derived masses of 37$^{+36}_{-19}$~\mjup\ for the outer companion in the
HD~38529 system, and of 34$\pm$12~\mjup\ for the outer companion in the
HD~168443 system, based on {\it Hipparcos} astrometry. This established the brown
dwarf nature of both objects. Most recently, \cite{sozzetti10} have followed that
same approach to derive masses for the two brown dwarf candidates
orbiting HD~131664 and HD~43848, respectively. With a mass of $23^{+26}_{-5}$~\mjup,
the companion to HD~131664 is indeed a brown dwarf, while the companion to
HD~43848 turned out to be stellar with a mass of $120^{+167}_{-43}$~\mjup.

Using the Fine Guidance Sensors (FGS) of the {\it Hubble} Space Telescope (HST),
an astrometric precision somewhat better than in the original {\it Hipparcos} Catalogue
can be achieved. Single measurement accuracies of around 1~mas and parallaxes as 
accurate as about 0.2~mas have been determined (\cite{benedict07,benedict09})
with HST; the most accurate parallaxes in the original {\it Hipparcos} Catalogue are
around 0.4~mas (for a few bright stars).
An upper mass limit of about 30~\mjup, at a 
confidence level of 99.73\%, was derived by \cite{mcgrath02} for the companion 
to 55~Cnc~b, based on HST/FGS data. Benedict et al.\ (\cite{benedict02}) reported 
the first
astrometrically determined mass of an extrasolar planet. They determined the
mass of the outermost planet orbiting GJ~876 as 1.89$\pm$0.34~\mjup, 
at 68.3\% confidence, based also on HST/FGS data. For the companion to 
$\epsilon$~Eri, \cite{benedict06} obtained a mass of 1.55$\pm$0.24~\mjup,
again at 68.3\% confidence and based on HST/FGS data. For the planet
candidate around HD~33636, \cite{bean07} derived a mass of 0.14$\pm$0.01~\msun
with the same method, implying that the companion is a low-mass star and not
a planet or a brown dwarf. 

Most recently, two brown dwarfs were confirmed with HST astrometry:
HD~136118~b has a mass of 42$^{+11}_{-18}$~\mjup\ (Martioli et
al.\ \cite{martioli10}), and  HD~38529~c has a mass of
17.6$^{+1.5}_{-1.2}$~\mjup\ (Benedict et al.\ \cite{benedict10}). In the 
$\upsilon$~And system, the inclinations of two companions could be measured,
which not only allowed for the determination of their masses 
(13.98$^{+2.3}_{-5.3}$~\mjup\ for $\upsilon$~And~c and
10.25$^{0.7}_{-3.3}$~\mjup\ for $\upsilon$~And~d, McArthur \cite{mcarthur10}), but
also the mutual inclination could be shown to be 29.9$\pm$1\degr. 
This is the first such measurement and shows the potential of astrometry for
the measurement of the 3-dimensional orbit geometry in multiple systems.

The median precision of positions and parallaxes in the
original version of the {\it Hipparcos} Catalogue is just better than 1~mas,
which is rather good, but still not good enough to detect those typical 
planetary companions which have been identified by radial-velocity surveys.

However, a new reduction of the raw {\it Hipparcos} data has been presented
by van Leeuwen (\cite{vanleeuwen07}). Through an improved attitude modeling,
systematic errors which dominated the error budget for the brighter stars in
particular were much reduced, by up to a factor of four compared to the original
version of the catalog. The formal error on the most precise parallaxes in the
original version of the {\it Hipparcos} Catalogue is around 0.4~mas, and around
0.1~mas in the new reduction presented by van Leeuwen (\cite{vanleeuwen07}).
The new reduction has been clearly shown to be superior
to the old reduction of the data in van Leeuwen (\cite{vanleeuwen07b}).
The smaller formal errors, in particular for the bright stars, greatly 
improve the prospect of finding astrometric signatures of planets and brown 
dwarfs in the data. 

In this paper we take a new look at the {\it Hipparcos} intermediate astrometric
data, based on the new reduction of the {\it Hipparcos} raw data by van Leeuwen
(\cite{vanleeuwen07}), for a large number of stars with planetary
or brown dwarf candidates from radial velocity surveys. 
With the improved astrometric accuracy, it might
be possible to detect a companion or place a tighter limit on its mass than
was possible before.

The outline of this paper is as follows: in Section~\ref{data}, 
we describe the various input data for our study, including the stellar sample
with known planetary and brown dwarf companion candidates as well as 
the astrometric data from {\it Hipparcos}. In Section~\ref{fitting}, we explain how
we fitted astrometric orbits to the new {\it Hipparcos} intermediate astrometric data.
Results are presented in Section~4 (upper mass limits for the companions to all 
examined stars) and in Section~5 (a few stars for which the astrometric orbit
could be detected). 
We conclude the paper with notes on individual stars in Section~6 and
a summary and discussion in Section~\ref{disc}. In the appendix, we show
how to calculate the astrometric signature and orientation of the apparent
orbit from the true orbit, taking projection effects into account.

\section{Data}
\label{data}

\subsection{Stellar Sample}

In an effort to be as exhaustive as possible, we put together a sample of all
known planetary and brown dwarf companions to {\it Hipparcos} stars
detected via radial velocities. We started with the list of planetary
companions compiled by \cite{butler06a}. We added those stars with substellar
companion candidates which were detected after 2006, as well as stars with
brown dwarf companions which were not included in the list by Butler. We
removed stars which were either not in the {\it Hipparcos} Catalogue or for which no
orbital elements were available, and we updated the orbital elements for those
stars for which new solutions were published in the meantime. 

We calculated periastron times for a 
few stars for which those were not given in the original table, namely transiting
planets with circular orbits\footnote{Formally, the periastron is not defined for
companions in circular orbits, but one can extend the usual definition by 
setting the longitude of the periastron to 0\degr, so that the periastron time
will refer to the time when the observed stellar radial velocity curve reaches 
its maximum. For circular orbits, this occurs exactly a quarter of a period before 
(hypothetical) mid transit time.} (HD~189733)
and planetary systems with significant interaction
between the companions (HD~82943, HD~202206 and GJ~876). 
For the latter, only osculating elements can be given, and
periastron times were calculated close to the epoch to which the elements referred.
For all three stars this was about a decade later than the {\it Hipparcos} epoch of J1991.25,
so that the orbital elements available might not be representative for the time
at which the {\it Hipparcos} measurements occurred. 

Furthermore, we assume that the longitude of periastron which is given for
spectroscopically detected extrasolar planets is actually the one pertaining to
the star, since this is the component observed (we verified this for a few
examples). The longitude of periastron of the two components differ by
180\degr; the distinction is important for the combination with positional data.

Our final list comprises 258~stars
with 310~substellar companions and is current as of April 2010. All stars
which were examined are listed in Table\ref{rvobs}, along with one or more 
references to the orbital elements. 

\subsection{{\it Hipparcos} intermediate astrometric data}
\label{newhip}

The new reduction of the {\it Hipparcos} Catalogue by van Leeuwen does not only include
a new estimate of the standard five astrometric parameters (mean positions, proper 
motions and parallax) for every star, but also the improved individual measurements,
the so-called intermediate astrometric data.

In contrast to the original solution, the given abscissa data are not averaged over all 
observations within an orbit in the van Leeuwen version. Rather, one abscissa residual 
is given per field transit, which increases the average number of individual abscissae
available for an object. The noise level of the new abscissae
is, after averaging, up to a factor of four smaller than before.
The errors of the averaged abscissae (per epoch accuracies) in the new
version of the {\it Hipparcos} Catalogue range from better than 0.7~mas for a few really 
bright stars up to around 10~mas for the faintest stars; most stars have abscissa
errors between 1.5 and 5~mas (van Leeuwen \cite{vanleeuwen07b}).

Everything else is very similar as before, although instead of the partial derivative
of each abscissa residual with respect to the five standard astrometric parameters,
the new version gives the instantaneous scan orientation and the parallax factor
instead. But this is just a different parameterization of the same information;
all relevant quantities can be derived from that, as explained in
van~Leeuwen (\cite{vanleeuwen07}).

\section{Orbit Fitting}
\label{fitting}

\subsection{Method}

We have fitted astrometric orbits to the new {\it Hipparcos} abscissa residuals for 
all stars listed in Table~\ref{rvobs},
simultaneously with corrections to the five standard astrometric parameters,
via a standard least squares minimization technique.
The only two orbital parameters fitted for were the inclination and the
ascending node; all other five orbital parameters (period, periastron time,
eccentricity, longitude of periastron and mass function) were kept fixed at the
literature values found via fits to the radial velocity data of each star.

If the star had an orbital solution in the {\it Hipparcos} Catalogue with a period
of the same order of magnitude as the spectroscopic one, we removed the astrometric
signature of the orbit from the abscissa data using the astrometric orbital
elements, and then fitted for the full orbit again using spectroscopic values
as input. In other words, we did not fit for corrections to the orbit as applied
in the {\it Hipparcos} Catalogue, but for the full orbit from scratch (this step is only
necessary for the abscissa data provided by van Leeuwen (\cite{vanleeuwen07}), 
since in the original {\it Hipparcos} Catalogue the abscissae always corresponded to 
a single star solution, even if an orbit was provided). 
This applies to three stars in our sample: HD~110833 (HIP~62145),
$\rho$~CrB (HIP~78459), and HD~217580 (HIP~113718). Likewise, we removed
the acceleration terms from the {\it Hipparcos} abscissa values if the solution 
was a 7 or 9 parameter solution before fitting for the astrometric orbit.
This applies to the following five stars: HD~43848 (HIP~29804), 55~Cnc (HIP~43587),
HD~81040 (HIP~46076), $\gamma^1$~Leo~A (HIP~50583), and HD~195019 (HIP~100970).
Please also note that the version of the van Leeuwen catalog which is available
in VizieR\footnote{\tt http://vizier.u-strasbg.fr/viz-bin/VizieR} is different
from the one provided together with the book from van Leeuwen (\cite{vanleeuwen07}).
E.g., 55~Cnc (HIP~45387) has a 7 parameter acceleration solution in the book version,
but a stochastic solution in the online version. Since the abscissae are only
available based on the book version, this is what we used here.

For stars with more than one substellar companion candidate, we did not fit
for these multiple companions simultaneously, but individually. This should be
a reasonable approach, since in the vast majority of the cases only one of the
companions will dominate the astrometric signal of the system. This is
especially true for systems detected via radial velocities, since for those all
companions tend to have similar radial velocity signals (i.e.\ the more massive
companions will be located further out). The astrometry would then be dominated
by that massive outer companion, whereas less massive inner companions have a
much smaller astrometric signal not detectable in the {\it Hipparcos} data.

We explicitly used all five spectroscopic orbital parameters in the 
fitting process and kept them fixed. The radial velocity signal is
in all the cases much more significant than the astrometric signal, so that
orbital parameters derived from the radial velocities should also be more
precise and accurate than those derived from the astrometry. Other authors
have chosen to disregard some of the spectroscopic orbital parameters, e.g.\
the radial velocity semi-amplitude (Mazeh et al.\ \cite{mazeh99}) or the 
longitude of the periastron (Pourbaix \& Jorissen \cite{pourbaix00}) 
and then later compared the astrometrically derived
value to the original, spectroscopic one, as a consistency check. 
However, we prefer not to disregard any spectroscopic
information in our astrometric fits, since this would likely compromise
the accuracy of the orbital parameters which we are most interested in and
not obtainable otherwise, the inclination and the ascending node. 

Likewise, we did not take the approach of fitting for the semi-major axis of 
the astrometric orbit
and only later linking this to the spectroscopic values via the parallax,
but applied all those constraints simultaneously and implicitly in the 
fitting process, which is a more direct approach and should yield
the highest accuracy in the inclination and ascending node.
We note that the criterion by Pourbaix \& Jorissen (\cite{pourbaix00}) 
which is often used to
link spectroscopic and astrometric quantities is not exactly correct,
as no allowance is made for projection effects. As detailed in the appendix,
the semi-major axis of the apparent orbit is possibly smaller than the 
semi-major axis of the true orbit, especially for eccentric orbits with
small inclinations.

Also, we did not compare the \chisq\ values of the standard solution (five
parameters) and the one including the astrometric orbit (two additional orbital
parameters) in an F test, as was done by Pourbaix \& Jorissen (\cite{pourbaix00}) 
to decide whether the
astrometric orbit was detected in the {\it Hipparcos} data. The reason is that we do
not want to evaluate whether the standard model or the orbital model is the
better one since we assume that the existence of a companion has been
established already by the radial velocity data. The question we would like to
ask here is how well the astrometric orbit can be constrained with the
{\it Hipparcos} abscissa data, and the most suitable criterion for this kind of
question is the joint confidence region allowed for the two orbital parameters. 
A small
confidence region, compared to the total allowed parameter range, indicates a
detection of the orbit, whereas a large confidence region would indicate no
real detection. In order to be conservative, we mostly use 3$\sigma$ confidence
regions, corresponding to a probability of 99.73\% that the true value falls
within this parameter interval.

In principle it would also be possible to fit for all orbital elements 
simultaneously using radial velocities as well as astrometry.
\cite{wright09} have developed an efficient algorithm which can fit several orbital 
companions to astrometric and radial velocity data jointly. Inclination
and ascending node will always be derived from the astrometric data alone,
since radial velocities are not sensitive to those parameters. The advantage
of jointly fitting both kinds of data are better constraints on the
five spectroscopic orbital parameters which now come from the two different
data sets, as well as explicit covariances between all orbital parameters. 
However, since in virtually all the
cases investigated here the companion signature is much more significantly
detected in radial velocities than in astrometry (if at all), the weighting
of the astrometric data in the combined fit would be much lower than that of
the radial velocity data, and in the end the astrometry would not influence
the values of the spectroscopic parameters. Therefore no attempt has been
made here to combine precise radial velocities with {\it Hipparcos} astrometry.
This situation should change once more accurate astrometry becomes available.

\subsection{$\epsilon$~Eri}

\begin{figure*}
\resizebox{\hsize}{!}{\includegraphics{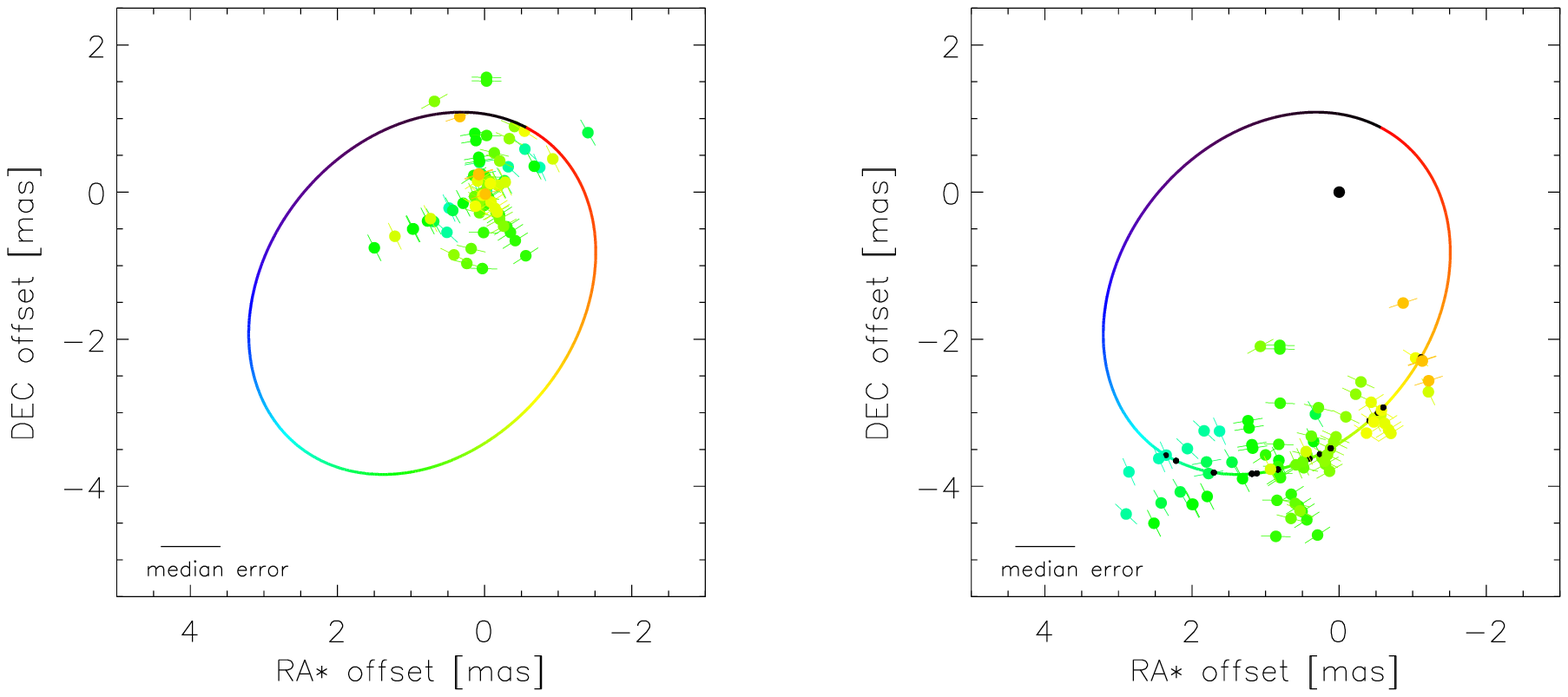}}\\[4ex]
\resizebox{\hsize}{!}{\includegraphics{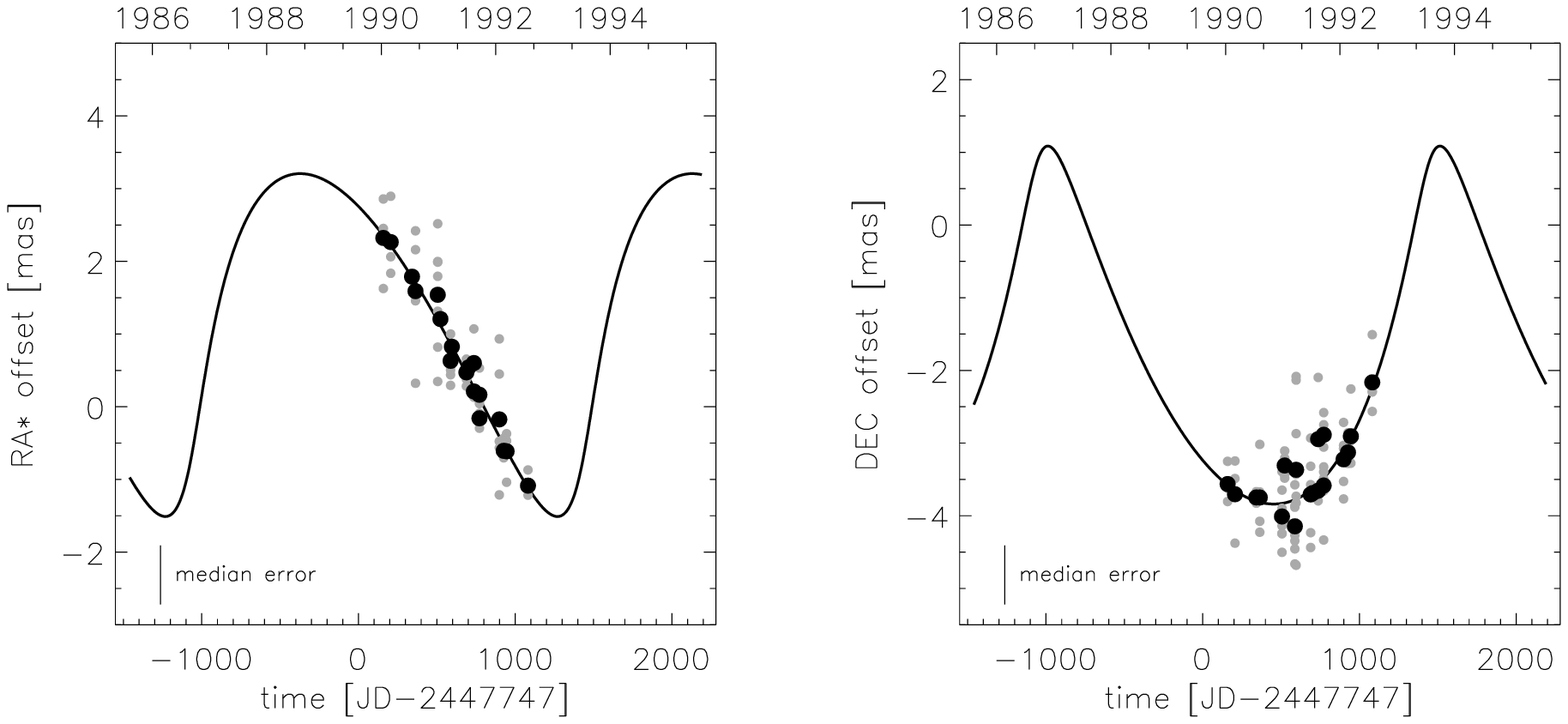}}
\caption{Illustration of abscissa residuals from {\it Hipparcos} along with the
astrometric orbit as fitted for $\epsilon$~Eri. The two panels in the top row
show the abscissa residuals (colored dots) with respect to the standard astrometric 
model without orbiting companion (upper left) and with respect to a model (solid line) 
which includes the companion (upper right). The colored lines indicate the direction
which was not measured and along which the dots are allowed to slide in the
adjustment process; the actual abscissa measurement is perpendicular to that line.
Color indicates orbital phase. The two panels in the bottom row show only one
dimension each as a function of time. The solid line is the orbital model as in 
the upper panels, while the small grey dots are the individual abscissa residuals.
The big black dots are averages of abscissa residuals taken very closely in time.
The information about the orientation of the abscissa residuals is missing in the
panels in the bottom row.
}
\label{epserifig}
\end{figure*}

An example for such an orbital fit is given in Fig.~\ref{epserifig} for the case
of $\epsilon$~Eri. The panels in the top row illustrate the 1-dimensional 
{\it Hipparcos} abscissa residuals (colored dots) in a 2-dimensional plot of the sky. 
The lines running through the colored dots indicate the direction which was not measured;
the dots (measurements) are thus allowed to slide on that line.
The actual measurement is perpendicular to the indicated line.
The upper left panel shows the abscissa residuals with respect
to the standard solution as given in the catalog, i.e.\ after fitting for the
five standard astrometric parameters, while the panel on the upper right shows the 
abscissa residuals for the case where the orbiting companion has been taken into account.
Thus, in the left panel the abscissa residuals are referred to the mean position of
the star at (0,0) in that diagram, while in the right panel the abscissa residuals
refer to the corresponding points on the orbit which are indicated by small black dots.
The solid line illustrates the astrometric orbit of the primary star as seen on the sky.
The color indicates orbital phase; an abscissa residual of a given color refers to the
point on the orbit with the same color. The time when a measurement was taken is an
additional constraint in the fit, and the color coding is an attempt to visualize that
additional constraint. 

Alternatively, the two panels in the bottom row each show one
dimension only as a function of time. The solid lines indicate the same orbit as
in the panels in the top row. The smaller grey points indicate the individual 
abscissa residuals (for the case where the orbital model has been taken into account 
as in the upper right panel), while the bigger black dots show the mean of all 
abscissa residuals taken very closely together in time. 
This is for illustration purposes only; the fit
was done using the individual, not the averaged abscissa residuals. 
When looking at the two panels in the bottom row please note that again not all 
constraints could be visualized at the same time; in these illustrations the information
about the orientation of the abscissae is lost. The measurements as indicated
in right ascension and declination are not static; due to the one-dimensional nature
of the abscissa residuals the values plotted in the lower panels can change in the
adjustment process. 

One important complication which affects some of our orbital fits is immediately
apparent: the {\it Hipparcos} measurements do not cover the full orbital phase range, 
but less
than half of that for the 6.9~year period of $\epsilon$~Eri~b. We will come back to
that point when we discuss the $\epsilon$~Eri system in more detail in Section~6.

\subsection{Verification}
For verification purposes, we attempted to reproduce the inclinations and
ascending nodes of spectroscopic binaries included in the {\it Hipparcos} Catalogue.
There are 235 spectroscopic binaries for which an orbit is listed in
the original {\it Hipparcos} Catalogue, and for 194 of those, 
the inclination and ascending node are actually obtained from a fit to the {\it Hipparcos} 
data (and not taken from some other reference).
For those 194~stars we tried to reproduce the fitted inclinations and ascending nodes,
using all other orbital elements as fixed input values (even if they were
fitted for in {\it Hipparcos}), so that our approach resembles most closely the
one which we are following with substellar companions detected with the
radial velocity method. We used the original version of the {\it Hipparcos} 
Catalogue from 1997, because here the intermediate astrometric data correspond
to the single star solution and because the orbital parameters are available
electronically, which both helps in the fitting process.

%In 39 cases our fit did not converge, and we disregarded those stars.
%For the remaining 155~stars, 
We derived inclinations and ascending nodes with errors for all 194 systems
and compared these values to the ones listed in the original {\it Hipparcos}
Catalogue. In 78\% of the cases, the two solutions agreed to within
0.3~$\sigma$, and in 97\% of the cases to within 1~$\sigma$ (although we 
note that the errors on the two angles can sometimes become rather large).
Still, we consider this a very satisfactory result, and are thus
confident that our method works correctly.

\section{Upper Mass Limits for Planets and Brown Dwarfs}
\label{masslim}

\addtocounter{table}{1}  % for the long table no. 1 at the end

While radial velocities provide \mtwosini\ (where $m_2$ is the companion mass and
$i$ is the inclination), a lower limit for the companion
mass, astrometry can provide an upper mass limit for the companion. This is
true even for stars where the astrometric signal of the companion
is too small to be detectable, since inclinations approaching 0\degr\ or
180\degr\ (face-on orbits) would yield companions which are so massive at
some point that they would show up in the astrometry. As the inclination of the
orbit approaches a face-on configuration, there is always a limit at which the
inclination is not compatible any more with the radial velocities and the astrometry.
This means that even for companions which are not detected in the astrometry,
there is usually still a (possibly weak) constraint on the inclination and thus
on the mass of the companion.

In order to derive upper mass limits, we fitted astrometric orbits to the
{\it Hipparcos} data for all stars on our list. The results are given in Table~\ref{rvobs}. 
The first two columns give the usual designation and the
{\it Hipparcos} number, respectively. The following column indicates if more information
on a particular star can be found in Section~\ref{notes}.
The reference column gives a numerical code to
the reference from which the spectroscopic orbital elements were taken; the
code is explained at the end of the table. The period column gives the 
period in days according to the reference in the previous column; it is always
derived from radial velocities. The following column gives the
minimum mass \mtwosini\ which corresponds to the orbital elements from the
given reference; the star mass used in the conversion is also taken from the
same reference. The last four columns give the result
of our astrometric orbit fitting. The first three of those give the minimum 
and the maximum inclination corresponding to the 3$\sigma$ confidence interval, 
$i_{\mbox{\scriptsize min}}$ and $i_{\mbox{\scriptsize max}}$, and the 
resulting 3$\sigma$ upper mass limit for the companion. In a few cases, 
an inclination of 90\degr\ could be excluded in the fitting. If this is the case,
a new, more stringent lower mass limit could be derived, too, which is denoted
as $m_{\mbox{\scriptsize 2,min}}$ and given in the last column, if applicable.
In some cases, especially when an inclination of 90\degr\ is not part of the
inclination confidence region, there are two minima in the $\chi^2$ map, and the 
confidence region for the second minimum is given in a second line for that
star. The table is sorted according to {\it Hipparcos} number (or right ascension,
respectively). 

%Note that upper mass limits can be derived for all companions, irrespective of
%whether the astrometric signature of the companion is actually detected in the
%{\it Hipparcos} data or not, as explained above (with the exception of a few stars for
%which the astrometric orbit fit did not converge; see below).

For nine companions, the derived upper mass limit lies in the planetary mass
regime, and thus unambiguously proves for the first time for most of them 
that these companions are really of planetary mass and not just planet candidates.
The confirmed planets are $\beta$~Gem~b (Pollux~b), \epseri~b, $\epsilon$~Ret~b, 
$\mu$~Ara~b, $\upsilon$~And~c and d, 47 UMa~b, HD~10647~b, and HD~147513~b. 
For $\epsilon$~Eri~b and $\upsilon$~And~c and d, the planetary nature was demonstrated
already by HST astrometry, and for $\upsilon$~And~d also by {\it Hipparcos} astrometry;
see Sections~\ref{epsiloneridani} and \ref{upsand}. 
For 47~UMa~b, an upper mass limit in the brown dwarf regime at 90\% confidence
was obatained before, based on the original {\it Hipparcos} data, but the planetary
nature could not be demonstrated unequivocally; see Section~\ref{47uma} for more
details.

For a further 75
companions the derived upper mass limit lies in the brown dwarf mass regime and
thus confirms at least the substellar nature of these companions. Two of those
(HD~137510~b and HD~168443~c)
have minimum masses also in the brown dwarf mass regime. These companions are
established brown dwarfs now instead of just brown dwarf candidates.
HD~168443~c was already confirmed to be a brown dwarf based on the original
{\it Hipparcos} astrometry; see Section~\ref{hd38529}.

\begin{figure}
\resizebox{\hsize}{!}{\includegraphics{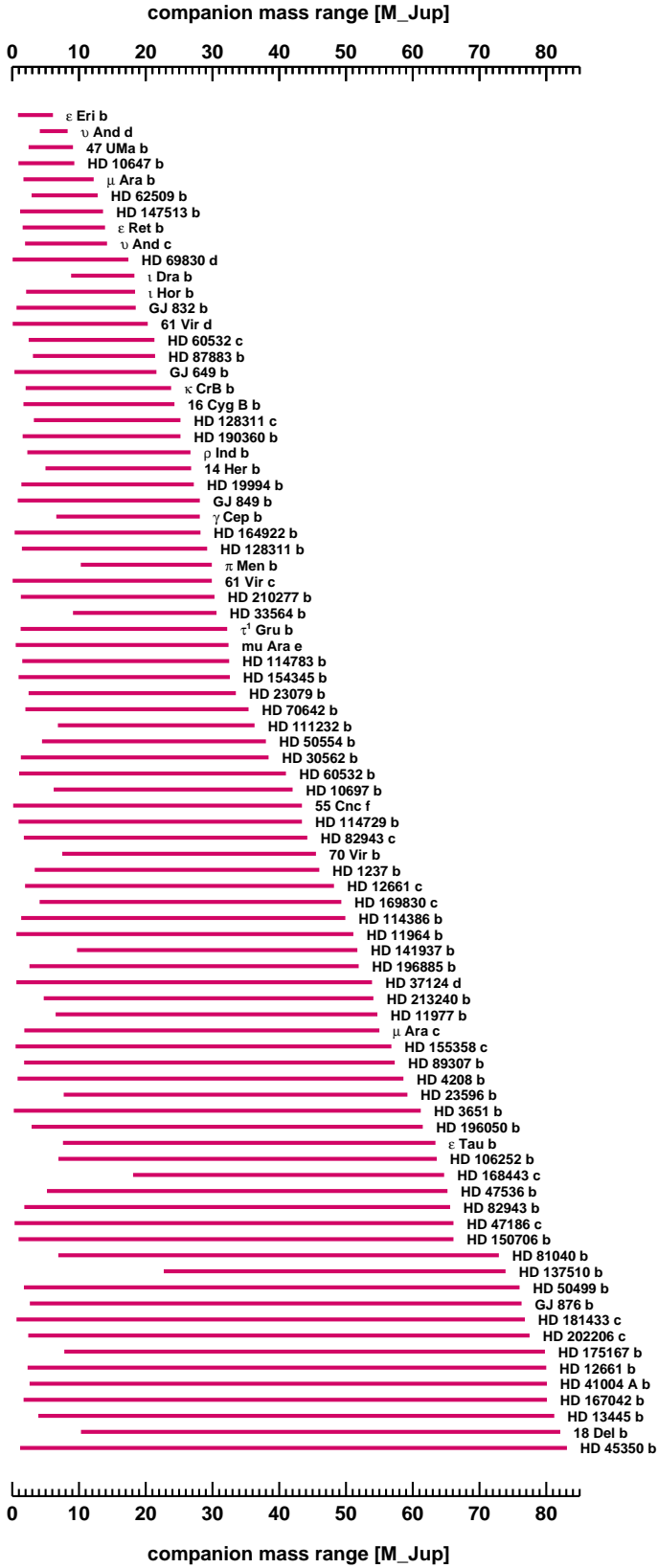}}
\caption{Illustration of the allowed 3$\sigma$ range of the companion mass for
all stars from Table~\ref{rvobs} for which the derived upper mass limit
from {\it Hipparcos} corresponds to less than 85~\mjup, i.e.\ for all 84~stars
for which the substellar nature could be confirmed by {\it Hipparcos}.
The lower mass limit usually corresponds to the lower mass limit
\mtwosini\ from the radial velocity fit, except for the few cases
where the astrometry yields a tighter lower mass limit by excluding
the inclination of 90\degr\ from the 3$\sigma$ confidence region
(HD~87883~b, HD~114783~b and $\gamma$~Cep~b).
The upper mass limit always comes from the {\it Hipparcos} astrometry, 
according to Table~\ref{rvobs}.}
\label{massfig}
\end{figure}

Fig.~\ref{massfig} illustrates the results for those 84 companions from
Table\ref{rvobs} for which the substellar nature could be established. It can
be seen that for many companions the allowed mass range between the lower mass
limit from radial velocities and the upper mass limit from astrometry is still
rather large. However, for a few of the companions this range is reasonably small
(see e.g.\ $\epsilon$~Eri~b, $\upsilon$~And~d or $\iota$~Dra~b),
and one might even speak of a determination of the real companion mass
rather than just placing a lower and upper limit on the mass, although there is
a continuous transition between the two. In Section~\ref{orbits} we will take a
closer look at those systems for which one might actually speak of a detection
of the astrometric orbit, rather than just a constraint on the inclination and 
thus the mass. 

\subsection{Limitations}

For some stars
the formally derived 3$\sigma$ upper mass limit of the companion is larger than 
5\,\msun, which
is not useful anymore. In fact, for astrometric upper mass limits larger than 
about 0.1-0.5\,\msun\ there might be more useful upper mass limits based on 
photometry and/or spectroscopy. A stellar companion that massive would show up
eventually in such data (see e.g.\ K\"urster, Endl \& Reffert \cite{kuerster08}). 
It has not been attempted to
derive those other upper mass limits here; the values given correspond to the
limits set by astrometry, and only under the assumption that the companion 
does not contribute a significant fraction to the total flux in the {\it Hipparcos}
passband. If the companion
was bright enough to affect the photocenter of the system, our method would fail 
since we assume that the photocenter is identical to the primary component in the system,
which we assume to be the {\it Hipparcos} star. If both components contribute significantly 
to the total flux, the observed orbit of the photocenter around the center of
mass of the system depends on the difference between flux ratio and mass ratio of
the two components. In order to model such a system, 
one would need to make additional assumptions about the components, which is
beyond the scope of this paper. 

Similarly, there are a number of stars for which the 3$\sigma$ confidence interval
in inclination extends from 0.1 to 179.9\degr. This is often the case for components
with very small minimum masses and/or very small periods derived from Doppler spectroscopy,
for which the astrometric signal is very small even for small inclinations. For those systems 
the derivation of the confidence interval in inclination is sometimes complicated by the fact that
the fitting process does not converge for all possible inclinations, as well as by the limited
resolution of our $\chi^2$ map in the interval between 0 and 0.1\degr\ or 179.9\degr\ and
180\degr, respectively. Of course it would be possible to increase the resolution for
those inclination intervals (we already used higher resolution for the intervals from 0 to 1\degr\ 
and from 179\degr\ to 180\degr\ than for the rest of the inclination range), but the results
would not be very meaningful since the method is just not suited very well for companions with
extremely small astrometric signatures. As a consequence, the derived upper mass limits for
those companions are not as accurate as those which correspond to inclinations which are not 
as close to 0 or 180\degr. 

For 67 companions no upper mass limit could be derived at all from the {\it Hipparcos}
astrometry. The reason is either that the astrometric fits did not converge 
(in 5000 iterations, which we set as a limit) over large parts of the
inclination and ascending node parameter space, or that the 3$\sigma$ confidence limits
in inclination are very close to 0\degr or 180\degr, respectively, as mentioned above.
In either case no meaningful confidence
limits in inclination or upper mass limits could be derived. 
For those stars we list the whole inclination interval
from 0\degr\ to 180\degr\ as 3$\sigma$ confidence limits, and the column giving
the 3$\sigma$ upper mass limit is left empty in Table~\ref{rvobs}. 54 of the 67~companions
without upper mass limits have periods smaller than 20~days, and 64 companions have periods
smaller than 50~days and thus very small expected astrometric signatures.

\section{Astrometric Orbits}
\label{orbits}

\begin{table*}
\caption{Best fit inclinations, ascending nodes and companion masses for stars where 
the astrometric orbit could be detected. $i_{\mbox{\scriptsize min}}$, 
$i_{\mbox{\scriptsize max}}$, and
$\Omega_{\mbox{\scriptsize min}}$, $\Omega_{\mbox{\scriptsize max}}$ correspond
to the limits of the 3$\sigma$ confidence region in both parameters jointly. The second row
for a star refers to a second local minimum where present. 
The following columns give the reduced $\chi^2$ value of the fit, the number of individual
{\it Hipparcos} abscissae available for that star, $n_{\mbox{\scriptsize obs}}$, and, for reference, 
the astrometric signature $\alpha$ for the fit.
Finally, the last four columns
give the minimum mass \mtwosini, derived from radial velocities, and the actual
mass of the companion $m_2$ derived here, with 3$\sigma$ confidence regions
($m_{2,\mbox{\scriptsize min}}$ and $m_{2,\mbox{\scriptsize max}}$). If the value in
the column indicating the lower limit of the 3$\sigma$ confidence region in mass is not given,
this means that an inclination of 90\degr\ cannot be excluded and thus that the lower mass limit
derived from radial velocities applies.
}
\label{orbittab}
\begin{minipage}{\textwidth}
\begin{tabular}{rrccccccccccccc}
\hline \hline\noalign{\smallskip}
 & & $i$ & $i_{\mbox{\scriptsize min}}$ & $i_{\mbox{\scriptsize max}}$ 
& $\Omega$ & $\Omega_{\mbox{\scriptsize min}}$ & $\Omega_{\mbox{\scriptsize max}}$ 
&  & 
& $\alpha$
& $m_2\sin i$ & $m_2$ 
& $m_{2,\mbox{\scriptsize min}}$ & $m_{2,\mbox{\scriptsize max}}$
\\
\multicolumn{1}{c}{\raisebox{1ex}[-1ex]{designation}}
& \multicolumn{1}{c}{\raisebox{1ex}[-1ex]{HIP no.}}
& [\degr] & [\degr] & [\degr] & [\degr] & [\degr] & [\degr]
& \multicolumn{1}{c}{\raisebox{1ex}[-1ex]{$\chi^2_{\mbox{\scriptsize red}}$}}
& \multicolumn{1}{c}{\raisebox{1ex}[-1ex]{$n_{\mbox{\scriptsize obs}}$}}
& [mas] & [\mjup] & [\mjup] & [\mjup] & [\mjup] \\[1.2ex]
\hline\noalign{\smallskip}
        HD 18445 b &  13769 & 147.6 &  31.3 & 160.2 & 285.7 & 252.2 &  20.4 &  1.34 & 110 &  4.32 &  44.    &   84.5 &       &  139.1 \\[0.8ex]
    BD $-$04 782 b &  19832 &  12.4 &  10.3 &  15.1 & 307.9 & 297.9 & 318.4 &  0.17 & 104 & 17.51 &  47.    &  261.6 & 207.4 &  329.2 \\[0.8ex]
        HD 43848 b &  29804 &  19.6 &  11.5 &  79.4 &  30.0 & 341.5 &  78.7 &  1.31 & 111 &  6.46 &   24.3\footnote{\cite{minniti09} 
give a value of 25~\mjup\ for \mtwosini; an iterative solution of the mass function however gives 24.3~\mjup.}
&   75.2 &  24.6 &  130.7 \\
                   &        & 158.3 &  98.2 & 167.1 &  99.2 &  48.5 & 152.3 &  1.32 &     &       &         &        &       &        \\[0.8ex]
           6 Lyn b &  31039 &   2.0 &   1.0 &  81.0 &  75.5 & 350.4 & 149.3 &  1.18 &  77 &  1.30 &   2.21  &   64.4 &   2.2 &  128.7 \\       
                   &        & 172.6 & 152.6 & 175.6 & 310.9 & 259.8 &  12.1 &  1.37 &     &       &         &        &       &        \\[0.8ex]
        HD 87883 b &  49699 &   8.5 &   4.8 &  57.7 & 254.9 & 182.2 & 316.5 &  0.79 &  77 &  2.75 &   1.78  &   12.1 &   2.1 &   21.4 \\       
                   &        & 168.9 & 133.7 & 173.9 &  96.4 &  40.7 & 164.0 &  0.86 &     &       &         &        &       &        \\[0.8ex]
$\gamma^1$ Leo A b &  50583 & 172.1 &  73.5 & 175.9 & 157.4 &  75.2 & 231.0 &  8.39 &  77 &  1.57 &    8.78 &   66.2 &   9.2 &  130.9 \\
                   &        &  13.6 &   7.5 &  59.3 & 359.9 & 298.4 &  66.7 &  9.50 &     &       &         &        &       &        \\[0.8ex]
       HD 106252 b &  59610 & 166.7 &   6.5 & 174.0 & 154.5 &  13.2 & 330.9 &  0.79 &  81 &  1.93 &   6.92  &   30.6 &       &   68.9 \\[0.8ex]
       HD 110833 b &  62145 &  10.4 &   7.7 &  16.8 & 120.9 &  96.4 & 149.4 &  2.03 & 171 &  5.97 &  17.    &  101.8 &  61.5 &  141.0 \\[0.8ex]
       HD 112758 b &  63366 &   8.9 &   6.5 &  15.2 & 150.5 & 126.9 & 174.9 &  1.77 &  81 &  4.85 &  34.    &  248.5 & 136.6 &  366.5 \\[0.8ex]
       HD 131664 b &  73408 & 167.1 & 149.3 & 171.9 & 320.8 & 276.4 &  10.0 &  0.76 & 154 &  4.13 &  18.15  &   85.2 &  36.3 &  139.7 \\[0.8ex]
     $\iota$ Dra b &  75458 &  69.9 &  26.6 & 141.8 & 182.9 &  10.7 & 324.9 &  0.78 & 137 &  0.24 &   8.82  &    9.4 &       &   19.8 \\[0.8ex]
      $\rho$ CrB b &  78459 &   0.4 &   0.4 &   0.7 & 266.4 & 242.3 & 290.6 &  2.67 & 173 &  1.96 &   1.093 &  169.7 & 100.1 &  199.6 \\[0.8ex]
       HD 156846 b &  84856 & 177.3 &   2.0 & 178.8 & 185.3 &  91.4 & 265.6 &  1.73 &  89 &  3.52 &  10.45  &  263.0 &       &  660.9 \\[0.8ex]
       HD 164427 b &  88531 &  12.2 &   7.1 &  44.7 & 337.5 & 294.7 &  29.9 &  1.15 &  92 &  2.27 &  46.4   &  244.2 &  66.8 &  458.5 \\[0.8ex]
       HD 168443 c &  89844 &  36.8 &  15.2 & 164.7 & 134.3 &  29.2 & 334.1 &  1.49 &  50 &  2.02 &  18.1   &   30.3 &       &   71.0 \\[0.8ex]
       HD 169822 b &  90355 & 175.1 & 172.5 & 176.1 & 249.8 & 236.1 & 267.5 &  0.79 & 165 &  8.49 &  27.2   &  388.7 & 237.4 &  527.9 \\[0.8ex]
       HD 184860 b &  96471 & 160.2 &  86.5 & 169.4 & 339.0 & 258.8 &  22.0 &  1.14 & 104 &  5.51 &  32.0   &   99.7 &       &  195.3 \\[0.8ex]
       HD 190228 b &  98714 &   4.5 &   2.4 & 174.1 &  71.0 & 219.4 & 164.2 &  1.08 & 254 &  1.63 &   5.93  &   76.8 &       &  147.2 \\[0.8ex]
       HD 217580 b & 113718 &  43.3 &  31.2 &  75.3 & 158.8 & 124.5 & 190.1 &  2.28 &  68 &  7.65 &  67.    &   99.9 &  69.0 &  135.8 \\[0.8ex]
    $\gamma$ Cep b & 116727 &   5.7 &   3.8 &  20.8 &  37.5 & 352.9 &  86.0 &  0.90 & 125 &  1.62 &   1.77  &   17.9 &   5.0 &   26.9 \\       
                   &        & 173.1 & 166.6 & 174.8 & 356.1 & 330.2 &  25.0 &  0.97 &     &       &         &        &       &        \\
\hline\noalign{\smallskip}
\end{tabular}
\end{minipage}
\end{table*}

\begin{figure*}
\resizebox{\hsize}{!}{\includegraphics{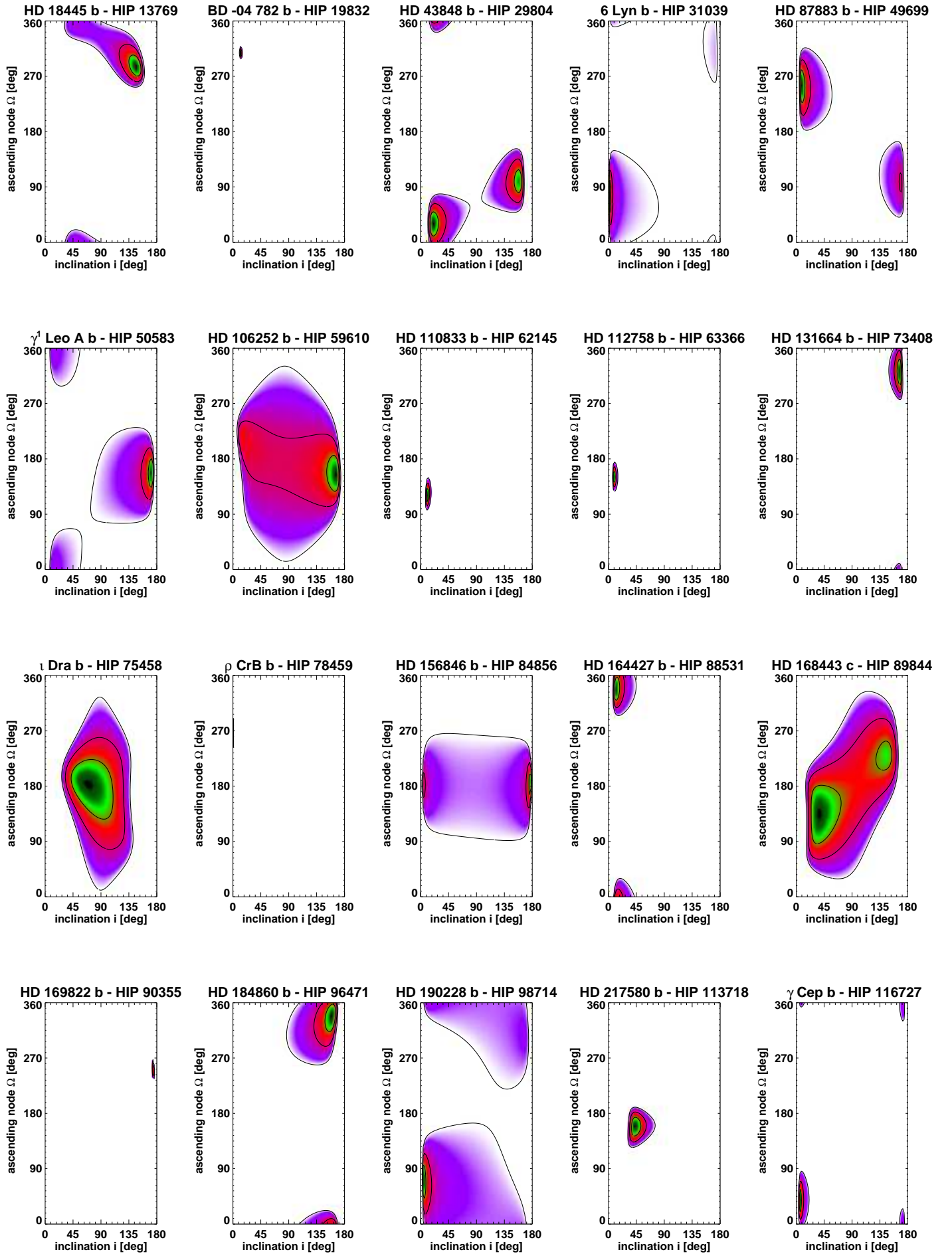}}
\caption{Illustration of the $\chi^2$ maps for those 20~stars for which the
astrometric orbit could be detected in the {\it Hipparcos} data, as a function of
inclination and ascending node. The contours show the $1\sigma$, $2\sigma$
and $3\sigma$ confidence regions in both parameters jointly. Please note that
for $\rho$~CrB~b the allowed range in inclination is so small that it is barely
visible in this representation; 
the corresponding data can be found in Table~\ref{orbittab}.}
\label{orbitfig}
\end{figure*}

For a few of the stars in Table~\ref{rvobs}, not only upper mass limits could be derived,
but also the astrometric orbit could be further constrained or fully determined.
While for most companions a limit on the inclination could be derived (even if the
lower limit is small), this is not necessarily true for the ascending node,
the only other orbital element which is not determined by the radial velocity fit.
This is due to the fact that an inclination approaching 0~degree will increase the size
of the orbit, until eventually the orbit would not be compatible any more with the
small scatter of the {\it Hipparcos} measurements. The ascending node however
describes the orientation of the orbit in space. Thus usually no or very weak
constraints can be placed on the ascending node if the orbit is not really
detected. However, for 20~systems of Table~\ref{rvobs}, at least a weak
constraint on the ascending node could be derived, indicating a (possibly weak)
detection of the astrometric orbit. In contrast to Section~\ref{masslim}, we
now use the confidence levels for two parameters (inclination and ascending
node) simultaneously, since we are now interested in the orbit, i.e.\ we want to 
derive constraints on both orbital parameters at the same time. Again, we used
the contours pertaining to a probability of 99.73\% ($3\sigma$). This will
always result in a less tight constraint on the inclination than given in
Table~\ref{rvobs}.

Those 20~systems with
detected astrometric orbits are shown in Fig.~\ref{orbitfig} and listed in 
Table~\ref{orbittab}. The table gives the best-fit inclinations and ascending nodes,
along with the 3$\sigma$ confidence limits. Note that the 3$\sigma$ confidence limits
(denoted as $i_{\mbox{\scriptsize min}}$, $i_{\mbox{\scriptsize max}}$ and 
$\Omega_{\mbox{\scriptsize min}}$, $\Omega_{\mbox{\scriptsize max}}$, respectively)
are now confidence regions in two parameters, inclination and ascending node, as opposed
to Table~\ref{rvobs}, where confidence limits in only one parameter (the inclination) were 
considered;
the corresponding confidence regions in Table~\ref{orbittab} are therefore slightly larger
than those in Table~\ref{rvobs}. Please also note that for some stars, two minima are visible
in the \chisq\ maps, often (but not always) with comparable local minimum \chisq\ values.
Those solutions correspond to orbits of about the same size (similar inclinations), but
with ascending nodes which differ by about 180\degr, so that the orbit has the opposite
orientation. For those stars affected, we give in Table~\ref{orbittab} the values
pertinent to the second minimum in a second line for that star. For stars for which
the confidence regions wrap around the 360\degr\ limit in the ascending node, continuing
at 0\degr, we list the higher value as $\Omega{\mbox{\scriptsize min}}$ and the lower one 
as $\Omega{\mbox{\scriptsize max}}$ in order to indicate the correct orientation of the 
confidence region. The following three columns give the reduced \chisq\ values,
the number $n_{\mbox{\scriptsize obs}}$ of individual 1-dim {\it Hipparcos} 
measurements (abscissae, or field transits) available for that star, and, 
for reference, the astrometric signature $\alpha$ for the fit.

\begin{table*}
\caption{Corrections (first rows) and final resulting values (second rows) of the
five standard astrometric parameters (mean positions, parallax, proper motions) for the
case where the astrometric modeling of the observations includes the effect of an orbiting
companion.}
\label{corrections}
\begin{tabular}{rrr@{.}lr@{.}lr@{.}lr@{.}lr@{.}l}
\hline \hline\noalign{\smallskip}
 &  & \multicolumn{2}{c}{$\Delta\alpha\star$ [mas]} & \multicolumn{2}{c}{$\Delta\delta$ [mas]} 
& \multicolumn{2}{c}{$\Delta\varpi$ [mas]}
& \multicolumn{2}{c}{$\Delta\mu_{\scriptsize \alpha\star}$ [\masa]} 
& \multicolumn{2}{c}{$\Delta\mu_{\scriptsize \delta}$ [\masa]} \\
\multicolumn{1}{c}{\raisebox{1ex}[-1ex]{designation}}
& \multicolumn{1}{c}{\raisebox{1ex}[-1ex]{HIP no.}}
& \multicolumn{2}{c}{$\alpha$ [rad]} & \multicolumn{2}{c}{$\delta$ [rad]} 
& \multicolumn{2}{c}{$\varpi$ [mas]} 
& \multicolumn{2}{c}{$\mu_{\scriptsize \alpha\star}$ [\masa]} & \multicolumn{2}{c}{$\mu_{\scriptsize \delta}$ [\masa]} \\[1.2ex]
\hline\noalign{\smallskip}
        HD 18445 b &  13769 &      $-$1&61 &         $-$2&95 &    1&65 &      0&63 &      0&99  \\       
                   &        & 0&7732664947 & $-$0&4358967094 &   40&00 &     14&45 &  $-$30&59  \\[0.8ex]
    BD $-$04 782 b &  19832 &      $-$8&83 &         $-$2&94 & $-$5&67 &   $-$4&43 &      8&50  \\       
                   &        & 1&1133364645 & $-$0&0771102807 &   42&40 &     85&93 &  $-$87&74  \\[0.8ex]
        HD 43848 b &  29804 &      $-$4&06 &         $-$4&63 & $-$1&40 &      3&73 &   $-$0&59  \\
                   &        & 1&6428834704 & $-$0&7074229223 &   25&03 &    125&77 &    197&73  \\[0.8ex]
           6 Lyn b &  31039 &         0&04 &         $-$0&12 &    0&12 &      0&66 &   $-$0&56  \\       
                   &        & 1&7051242326 &    1&0151438236 &   18&04 &  $-$29&50 & $-$339&24  \\[0.8ex]
        HD 87883 b &  49699 &      $-$0&14 &         $-$0&41 &    0&30 &      1&82 &   $-$0&00  \\       
                   &        & 2&6560409615 &    0&5976427335 &   55&23 &  $-$62&23 &  $-$60&51  \\[0.8ex]
$\gamma^1$ Leo A b &  50583 &         2&40 &            0&56 &    2&11 &   $-$1&89 &   $-$1&01  \\
                   &        & 2&7051266777 &    0&3463057931 &   27&11 &    302&52 & $-$155&30  \\[0.8ex]
       HD 106252 b &  59610 &      $-$0&96 &         $-$1&72 & $-$0&02 &      1&29 &      0&99  \\       
                   &        & 3&2004608067 &    0&1752714978 &   26&53 &     25&05 & $-$278&51  \\[0.8ex]
       HD 110833 b &  62145 &         2&04 &            7&04 & $-$0&66 &      0&54 &      0&18  \\
                   &        & 3&3346626142 &    0&9033780697 &   66&54 & $-$378&54 & $-$183&51  \\[0.8ex]
       HD 112758 b &  63366 &         0&63 &         $-$2&50 &    0&50 &   $-$1&55 &      2&05  \\       
                   &        & 3&3991779486 & $-$0&1716455024 &   48&36 & $-$826&79 &    198&20  \\[0.8ex]
       HD 131664 b &  73408 &      $-$0&14 &         $-$2&23 &    0&93 &   $-$1&54 &   $-$2&37  \\       
                   &        & 3&9274303384 & $-$1&2834350195 &   18&68 &     13&08 &     26&49  \\[0.8ex]
     $\iota$ Dra b &  75458 &      $-$0&14 &         $-$0&03 &    0&04 &   $-$0&04 &      0&04  \\       
                   &        & 4&0357673046 &    1&0291512587 &   32&27 &   $-$8&40 &     17&12  \\[0.8ex]
      $\rho$ CrB b &  78459 &      $-$0&27 &         $-$0&76 &    0&65 &      1&49 &      0&07  \\
                   &        & 4&1933570759 &    0&5812886759 &   58&67 & $-$195&15 & $-$772&93  \\[0.8ex]
       HD 156846 b &  84856 &         3&43 &         $-$3&83 & $-$0&94 &   $-$0&33 &      0&24  \\       
                   &        & 4&5403574027 & $-$0&3374314890 &   20&05 & $-$137&00 & $-$144&00  \\[0.8ex]
       HD 164427 b &  88531 &         0&38 &            1&51 & $-$0&56 &      0&08 &      0&21  \\       
                   &        & 4&7329564103 & $-$1&0333998216 &   25&76 & $-$199&22 &  $-$51&31  \\[0.8ex]
       HD 168443 c &  89844 &      $-$0&71 &            0&50 & $-$0&37 &      1&34 &   $-$0&10  \\       
                   &        & 4&7999453113 & $-$0&1674674410 &   26&35 &  $-$92&47 & $-$222&94  \\[0.8ex]
       HD 169822 b &  90355 &         8&35 &            4&24 & $-$3&31 &   $-$2&39 &   $-$2&24  \\       
                   &        & 4&8265774036 &    0&1532170533 &   31&32 & $-$196&50 & $-$462&44  \\[0.8ex]
       HD 184860 b &  96471 &      $-$2&71 &         $-$5&95 &    0&36 &      1&20 &      0&59  \\       
                   &        & 5&1345996462 & $-$0&1822607121 &   36&15 & $-$292&58 & $-$273&12  \\[0.8ex]
       HD 190228 b &  98714 &         0&41 &         $-$1&37 & $-$0&22 &   $-$1&16 &      0&30  \\       
                   &        & 5&2491288882 &    0&4940508070 &   16&03 &    104&05 &  $-$69&54  \\[0.8ex]
       HD 217580 b & 113718 &         2&39 &            3&99 &    6&29 &      1&19 &      1&85  \\
                   &        & 6&0294808555 & $-$0&0671642787 &   65&00 &    397&39 & $-$205&83  \\[0.8ex]
    $\gamma$ Cep b & 116727 &         0&27 &         $-$0&10 &    0&18 &      0&69 &      0&36  \\       
                   &        & 6&1930813876 &    1&3549334224 &   71&10 &  $-$47&27 &    126&95  \\[0.8ex]
\hline\noalign{\smallskip}
\end{tabular}
\end{table*}

The last four columns of Table~\ref{orbittab} give the resulting masses for the astrometric
solution. The minimum mass $m_2\sin i$ derived from the radial velocities is listed in the first
of the four mass columns. This is also the minimum mass for any of the
astrometric solutions, so it is given here for reference. It would correspond to an 
inclination of 90\degr. The next mass column labeled $m_2$ gives the mass which corresponds
to the best fit inclination. There is only one best fit mass for each system, so the second
line is empty even if there are two minima. The next to last column, labeled 
$m_{2,\mbox{\scriptsize min}}$, gives the lower limit for the 3$\sigma$ confidence region
in mass. If the inclination value of 90\degr\ is part of the 3$\sigma$ confidence region 
(see also Fig.~\ref{orbitfig} or the columns $i_{\mbox{\scriptsize min}}$ and 
$i_{\mbox{\scriptsize max}}$), then this value corresponds
exactly to the minimum secondary mass derived from radial velocities. In order to indicate that 
this is the case, no mass is given here if appropriate. It can be seen that for twelve systems
an inclination of 90\degr\ can be excluded with
99.73\% confidence; for all others, the minimum secondary mass derived from radial velocities,
$m_2\sin i$, is within the 3$\sigma$ mass confidence region and thus also the lower limit
for the confidence region of the mass.

Some orbits are rather well determined, even if using the conservative 99.73\% 
confidence regions. The inclinations of BD~--04~782~b, $\rho$~CrB~b and HD 169822~b could
be determined to better than 5\degr, while the inclinations of HD~110833~b and HD~112758~b
could be determined to better than 10\degr. 
For those systems, the ascending node is also rather well constrained. 
HD~168443~b could be confirmed as brown dwarf, even if taking the conservative 3$\sigma$ 
confidence levels into account. $\iota$~Dra~b
is most likely a high-mass planet, while HD~106252~b and $\gamma$~Cep~b are most likely 
brown dwarfs. The companions for which the inclination could be determined to better than
5\degr\ are all low-mass stars.

Altogether, using the best fit masses, we find two planets (HD~87883~b and $\iota$~Dra~b),
eight brown dwarfs and ten stars among our 20~systems
with astrometric orbits. Note that this is by no means representative for our whole
sample listed in Table~\ref{rvobs}, since it is much more likely to find the astrometric
signature of a star or a brown dwarf than the tiny signature of a planet in the 
{\it Hipparcos} data. 

For completeness, we also list the corrections to the five standard astrometric parameters
(mean positions at the {\it Hipparcos} epoch of J1991.25, parallax and proper motions)
in the new {\it Hipparcos} Catalogue and the new resulting values for these in Table~\ref{corrections}.
These astrometric parameters were fitted simultaneously with the orbital parameters, 
and the changes are a direct consequence of the new fitting model which now includes 
the astrometric orbit.
As usual, $\alpha\star$ denotes a value where the $\cos\delta$ has been factored in
already. Note that the positional corrections are given in milli-arcseconds, whereas the 
coordinates themselves are given in radians, just like in the {\it Hipparcos} Catalogue by
van Leeuwen (\cite{vanleeuwen07}). 
The corrections to the astrometric parameters are usually rather small,
typically of the order or smaller than 1~mas or 1~\masa, respectively.

\section{Notes on individual stars}
\label{notes}

\subsection{$\epsilon$~Eri (HIP~16537)}
\label{epsiloneridani}
The presence of a planet orbiting \epseri\ (HIP~16537, V=3.7~mag) had long been 
suspected (Walker et al.\ \cite{walker95}). In 2000, \cite{hatzes00} finally announced 
the secure
detection of \epseri~b, combining a variety of radial velocity data sets taken
by different groups. It orbits its star in about 6.9~years, and has a minimum
mass of $m\sin i=0.86$~\mjup. Because of its small distance from the Sun 
(3.2~pc) the expected minimum astrometric signature amounts to 1.9~mas.

Using {\it Hubble} Space Telescope (HST) data in combination with data from the
Multichannel Astrometric Photometer (MAP) as well as radial velocities, 
\cite{benedict06} fit for the astrometric orbit of \epseri\ due to 
\epseri~b. They derive an inclination of 30\fdg1$\pm$3\fdg8 and an
ascending node of 74\degr$\pm$7\degr, yielding a mass for the companion
of 1.55$\pm$0.24~\mjup.

This is in excellent agreement with the astrometric fit to the {\it Hipparcos} data 
derived in this paper. Our best fit values are an inclination of 
23\degr\ and an ascending node of 282\degr, with 1$\sigma$ errors of the 
order of 20\degr and a lower limit for the inclination of 15\degr (1$\sigma$)
and 8\fdg8 (3$\sigma$), respectively. We obtain a companion mass
of $2.4\pm1.1$~\mjup, slightly larger than the HST value.
A second minimum is found at an inclination of 158\degr\ and an ascending
node of 53\degr, which is very similar to the original orbit except that
ascending and descending nodes are exchanged. 
We do not consider the orbit detected
in the {\it Hipparcos} data according to our criterion chosen in Section~\ref{orbits},
because the 3$\sigma$ confidence region in inclination and ascending node
jointly encompasses the whole parameter range for the ascending node, and rather
count \epseri\ among the stars with derived upper mass limits, but not with
detected orbital signatures.

We have used the robust radial velocity fit from \cite{hatzes00} as an
input for our astrometric fit to the {\it Hipparcos} intermediate astrometric data.
Formally the fit from \cite{benedict06} would have
been the most precise one, but we preferred input parameters that came from a
spectroscopic fit only instead of one which already was based on combined 
spectroscopy and HST astrometry. The slight differences which we find in
companion masses are thus partly due to the use of different input data.

As mentioned already the phase coverage of the astrometric data is
not complete. This applies to both the HST/FGS as well as the {\it Hipparcos} data;
the HST/FGS data cover 2.9~years, while the {\it Hipparcos} data cover 2.5~years. This
is to be compared with the orbital period of 6.9~years. Unfortunately, neither
of the two astrometric data sets covers the phase of the orbit where the motion
in right ascension is reversed, which would be very helpful to distinguish
between orbital curvature and linear proper motion.

It is not straightforward to combine the {\it Hipparcos} data with the HST/FGS data 
from \cite{benedict06}, since the reference points for the astrometry
are different. In modeling the {\it Hipparcos} data, we have to simultaneously fit
for orbital as well as astrometric parameters (mean absolute positions
at the {\it Hipparcos} epoch of J1991.25, absolute proper motions and absolute
parallax). In contrast to that, the HST/FGS astrometry is relative to a 
frame of standard stars, although the relative parallax has been converted 
into an absolute one by determining the parallaxes of the reference frame 
stars.
In addition, the approach followed by \cite{benedict06} was to derive
proper motions and parallaxes from measurements over a much larger time
interval than their HST/FGS observations by including observations from 
the Multichannel Astrometric Photometer (MAP). If the {\it Hipparcos} and the HST/FGS
data were to be combined, one would have to homogenize the astrometric 
parameters (positions, proper motions, parallaxes) and fit for additional
zero-point offsets between the two datasets. However, since the zero-point
of the HST/FGS proper motions is unknown, it is not possible to combine
the two astrometric datasets in a rigorous way.

\subsection{$\upsilon$~And (HIP~7513)}
\label{upsand}
We derive rather tight constraints on the masses of the two outermost
companions in the three planet system: 
an upper mass limit of 8.3~\mjup\ (99.73\% confidence) for $\upsilon$~And~d, 
and an upper mass limit of 14.2~\mjup\ for $\upsilon$~And~c;
the minimum masses determined from radial velocities
are 4.13~\mjup\ ($\upsilon$~And~d) and 1.92~\mjup\ ($\upsilon$~And~c), respectively.
Our newly derived upper mass limits place a much tighter constraint
on the planet masses than was done before by Mazeh et al.\ (\cite{mazeh99}), 
who derived an upper mass limit of 19.6~\mjup\ at 95.4\% confidence for
$\upsilon$~And~d, using
the original version of the {\it Hipparcos} Catalogue. 

Most recently, rather accurate masses for those two companions in the 
$\upsilon$~And system were measured from HST astrometry (McArthur et al.\ \cite{mcarthur10}):
$\upsilon$~And~d has a mass of $10.25\pm^{+0.7}_{-3.3}$~\mjup, and 
$\upsilon$~And~c has a mass of $13.98\pm^{+2.3}_{-5.3}$~\mjup. The
corresponding best-fit inclinations are rather small ($7\fdg868\pm 1\fdg003$ for
$\upsilon$~And~d and $23\fdg758\pm 1\fdg316$ for $\upsilon$~And~c). These results
agree well with each other, although our upper mass limits are both smaller
than the measured HST masses. 

Since $\upsilon$~And is rather bright (V=4.1~mag),
the {\it Hipparcos} abscissae are not photon-noise dominated as for other stars,
but are of comparable accuracy as the HST data. The single measurement precision
is about 1~mas for both HST and {\it Hipparcos}. {\it Hipparcos} took data at 28 different 
epochs, covering 3.2~years, while the HST data refer to 13 different
epochs, covering approximately 5~years. The precision per epoch is higher for 
HST, since more single measurements are averaged. Resulting parallaxes are rather 
precise for both HST and {\it Hipparcos}, with formal errors of 0.10~mas for the 
HST parallax and 0.19~mas for the {\it Hipparcos} parallax, respectively. 

One should
note further that our approach of fitting one companion candidate at a time
to the astrometric data
is not ideal in the $\upsilon$~And system, where the two outermost companions both 
contribute to the observed astrometric signal. However, it is unlikely that a
combined fit of the two companions would lead to considerably higher upper mass
limits; in general one would expect smaller astrometric signatures and thus smaller
masses in a combined fit of the two companions as compared to a fit of both
components separately. Thus our upper mass limits should hold even in the
$\upsilon$~And system, although they might be too conservative.

\subsection{47~UMa (HIP~53721) and 70~Vir (HIP~65721)}
\label{47uma}
For the companion to 47~UMa, an upper mass limit of between 7 (best fit) and
22~\mjup\ (90\% confidence level) was derived by \cite{perryman96}, using the same
method as we employ here, but with the original version of the
{\it Hipparcos} Catalogue (ESA~\cite{esa97}).
The minimum companion mass from the RV solution is 2.45~\mjup.
We derive a tighter constraint on the mass, with an upper mass limit of 9.1~\mjup\ 
at 99.73\% confidence, which demonstrates that the companion is a planet.
The best fit value for the mass is 3.0~\mjup, and the 1$\sigma$ upper mass limit 
is 4.9~\mjup.

An upper mass limit of 38--65~\mjup\ was derived for the companion
to 70~Vir by \cite{perryman96}, where 38~\mjup\ corresponds
to the best fit, while 65~\mjup\ corresponds to a confidence level of 90\%.
We derive an upper mass limit of 45.5~\mjup, corresponding to a rather conservative 
confidence level of 99.73\%. Our 1$\sigma$ 
upper mass limit would be 26.8~\mjup, smaller than the best fit value of 
\cite{perryman96}.

For 51~Peg (HIP~113357), neither \cite{perryman96} nor we could derive a very
meaningful result; the upper mass limit derived is around 0.5~\msun\ in both
investigations.

Overall, our results are thus fully consistent with the results derived by
\cite{perryman96}. Our constraints on the mass are a bit tighter, just as 
expected for the improved {\it Hipparcos} data by van Leeuwen (\cite{vanleeuwen07}).

\subsection{HD~33636 (HIP 24205)}
Based on HST astrometry, \cite{bean07} find an inclination of $4.0\pm0.1$\degr\
and an ascending node of $125\fdg6\pm1\fdg6$. This translates into a companion
mass of 142$\pm$11~\mjup, making the companion a low-mass star.
Our {\it Hipparcos} solution is similar to the solution by \cite{bean07},
although our inclination is slightly larger and thus our conclusion on the
companion mass is slightly different.

We find a best-fit inclination of 11\fdg2 and an
ascending node of 151\fdg7 from a fit to the {\it Hipparcos} data.
The 1$\sigma$ confidence region in inclination
extends from 6\fdg2 to 39\fdg0 while the 3$\sigma$ confidence region
extends from 2\fdg9 to 176\fdg8. The corresponding upper mass limits 
are 90~\mjup\ (1$\sigma$) and 207~\mjup\ (3$\sigma$), respectively.
The confidence region in the ascending node extends over the whole
parameter range, so that we do not consider the astrometric orbit detected
in the {\it Hipparcos} data. The {\it Hipparcos} data are less precise than the HST
data for HD~33636. For comparison, the formal error on the parallax 
is 0.2~mas for the HST data, and 1.0~mas for the {\it Hipparcos} data.

Our 3$\sigma$ lower limit on the inclination and upper limit on the mass is
thus fully consistent with the conclusion of \cite{bean07} that the companion to
HD~33636 is actually a low-mass star, although our solution would
favor a brown dwarf companion.

However, one should note that the $\chi^2$ value for our fit, as well as in
the original {\it Hipparcos} data, is a little bit on the high side, indicating 
that the astrometric model applied might not be fully adequate.

\subsection{HD~136118 (HIP~74948)}
Martioli et al.\ (\cite{martioli10}) derive a mass of 
42$^{+11}_{-18}$~\mjup\ for the companion
to HD~136118 based on HST astrometry, confirming it to be a brown dwarf.
The best fit inclination and ascending
node are 163\fdg1$\pm$3\fdg0 and 285\degr$\pm$10\degr, respectively.
We do not detect the astrometric orbit, but derive an upper mass limit of 
95~\mjup\ (3$\sigma$ confidence limits), which also makes the companion most 
likely a brown dwarf (the minimum mass from radial velocities is about 12~\mjup). 
Our best fit values for inclination and ascending node are 152\degr\ and 294\degr,
respectively, which is in rather good agreement with the HST values. It seems
that {\it Hipparcos} has actually weakly detected the astrometric signature, although
it is below our conservative significance level to be called a detection.

\subsection{HD~38529 (HIP~27253) and HD~168443 (HIP~89844)}
\label{hd38529}
Both systems harbor two companions, of which the outer ones are most
likely brown dwarfs; the minimum mass from Doppler spectroscopy
is about 13~\mjup\ for HD~38529~c and about 18~\mjup\ for HD~168443~c.
We had fitted astrometric orbits to HD~38529 and HD~168443 using the original
version of the {\it Hipparcos} intermediate astrometric data already in 
Reffert \& Quirrenbach (\cite{reffert06b}). According to our convention
followed here to call an astrometric orbit detected if there is a constraint
on the ascending node using 3$\sigma$ confidence levels, the astrometric orbit
of HD~38529~c was not detected in the original version of the {\it Hipparcos} data, whereas
the astrometric orbit of HD~168443~c was detected. The same holds true for the 
analysis based on the new version of the {\it Hipparcos} data presented in this paper,
although the details on the fitted angles and corresponding upper mass limits differ somewhat.

For HD~38529~c, the older best fit values were 
160\degr$^{+7^{\circ}}_{-23^{\circ}}$ for the inclination
and 52\degr$^{+24^{\circ}}_{-23^{\circ}}$ for the ascending node. Here we derive a best 
fit inclination of
44\fdg1 (1$\sigma$ confidence region from 18\fdg6 to 161\fdg1) 
and a best fit ascending node of 54\fdg4. The ascending node is thus
in perfect agreement, whereas the inclinations differ somewhat. The best fit mass
was 37$^{+36}_{-19}$~\mjup based on the original {\it Hipparcos} data, and is now 
18.8$^{+22.5}_{-5.7}$~\mjup. 
However, our new solution is in perfect agreement with the HST astrometry 
which has become available for HD~38529~c in the meantime 
(Benedict et al.\ \cite{benedict10}):
best values are 48\fdg3$\pm$3\fdg7 for the inclination and 38\fdg2$\pm$7\fdg7 for
the ascending node, which implies a best fit mass of 17.6$^{+1.5}_{-1.2}$~\mjup.

For HD~168443~c, our best fit values based on the original version of the
{\it Hipparcos} Catalogue were 150\degr$^{+8^{\circ}}_{-20^{\circ}}$ for the inclination and 
19\degr$^{+21^{\circ}}_{-24^{\circ}}$ for the
ascending node. With the new version of the {\it Hipparcos} intermediate astrometric
data, we derive a best inclination of 36\fdg8 
(1$\sigma$ confidence region from 27\fdg4 to 54\fdg9)
and a best fit ascending node
of 134\fdg3, with a second minimum at an inclination of 145\fdg4 and an
ascending node of 230\fdg0 (in the original version, there was only one
minimum in the $\chi^2$ map). This implies a best fit mass of 
30.3$^{+9.4}_{-12.2}$~\mjup,
which compares favorably to the original value of 34$\pm$12~\mjup.
The 3$\sigma$ upper mass limit is about 65~\mjup, so that the companion is 
confirmed as a brown dwarf. This is also consistent with the analysis of
\cite{zucker01}, who derived an upper mass limit of about 80~\mjup\ using
the original version of the {\it Hipparcos} data.

\subsection{55~Cnc (HIP~43587) and GJ~876 (HIP~113020)}
The HST/FGS limits for both stars are tighter than the ones which could be
derived here from {\it Hipparcos} data. For 55~Cnc~b, \cite{mcgrath02} derive an
upper mass limit of 30~\mjup, whereas we derive 0.26\,\msun, both with
99.73\% confidence. For 55~Cnc~d, a preliminary analysis of HST data yields
an inclination of $53\degr\pm6\fdg8$ (\cite{mcarthur04}). 
It is mentioned that in any case the inclination
could not be smaller than 20\degr, even if considering the fact that the HST data
cover only a small part of the whole orbit. We do not find any evidence of
an astrometric orbit for any of the currently known five companions to 
55~Cnc. The minimum inclinations are between 0\fdg1 and 2\fdg8, which does not 
result in tight constraints on the masses.

One should note that while the original {\it Hipparcos}
Catalogue solved HIP~43587 as a single star, the new version by van Leeuwen
(\cite{vanleeuwen07}) adds accelerations in the proper motions to the astrometric
fit, but even with those additional parameters the fit quality in the new 
{\it Hipparcos} Catalogue is very poor. It does not improve considerably when we fit
an orbital model instead; the resulting reduced $\chi^2$ value is among the highest
found for all examined stars, and our solution should be treated with caution.
In particular, it is not meaningful to derive confidence regions when the underlying
fitting model does not represent the data adequately.

For GJ~876~b, Benedict et al.\ (\cite{benedict02}) obtain a mass of
1.89$\pm$0.34~\mjup. We do not detect the astrometric orbit but
derive an upper mass limit of 76~\mjup, at 99.73\% confidence.

\subsection{$\rho$~CrB (HIP~78459)}
The astrometric signature of the substellar companion candidate around 
$\rho$~CrB is formally detected, with a 3$\sigma$ confidence region in
inclination and ascending node which rejects a huge part of the available 
parameter space.

$\rho$~CrB was solved as a double star with an orbital solution in both
versions of the {\it Hipparcos} Catalogue, with a period of about 78~days and
a photocenter motion of about 2.3~mas. However, the orbital fits are not
very good. Interestingly, the period of $\rho$~CrB~b (39.8449~days, 
Noyes et al.\ \cite{noyes97}) is about half of the {\it Hipparcos} period, and it is 
noted by van Leeuwen (\cite{vanleeuwen07}) that in some cases a factor two
ambiguity could be present in the {\it Hipparcos} periods. So it might very well 
be possible that the signature of the companion to $\rho$~CrB (which is a low-mass
star according to our new orbital fit) was actually detected even in the 
original version of the {\it Hipparcos} Catalogue!

\subsection{HD~283750 (HIP~21482) and $\gamma^1$~Leo~A (HIP~50583)}
The astrometric fits for HD~283750 and $\gamma^1$~Leo~A carry somewhat 
large $\chi^2$ values, and are therefore to be treated with caution. 

HD~283750 is a spectroscopic binary with a 1.8~day period (Halbwachs et al.\ 
\cite{halbwachs00}). \cite{tokovinin06}
list a third component in the system which could be identified in the 2MASS catalog 
and which might be responsible for the larger than usual astrometric residuals.
The two close components of HD~283750 have probably formed as a close stellar binary
system, rather than as a primary star with a low-mass substellar companion forming in
the disk.

$\gamma^1$~Leo~A is a binary with a separation of about 4.6\arcsec; the primary is a
K~giant and the secondary is photometrically variable with a period of about 1.6~days 
according the notes in the {\it Hipparcos} Catalogue. Clearly, the astrometry could be
affected by the double star nature. We detected the astrometric orbit of the companion
of the primary, but we caution that the $\chi^2$ value is unusually large.
In the version of the {\it Hipparcos} Catalogue by van Leeuwen (\cite{vanleeuwen07}), 
$\gamma^1$~Leo~A is solved with a model including accelerations in the proper motions.

\subsection{HD~43848 (HIP~29804)}
\label{hd43848}

The brown dwarf candidate HD~43848~b with a minimum mass of around 25~\mjup\ was discovered 
by \cite{minniti09} via Doppler spectroscopy. We detect the astrometric orbit, and derive
a best fit inclination of 19\fdg6, corresponding to a mass of 75~\mjup, and
a best fit ascending node of 30\fdg0. The upper mass limit from 
the joint 3$\sigma$ confidence regions in inclination and ascending node is 131~\mjup.

\cite{sozzetti10} have performed a very similar analysis, but arrived at a slightly different 
conclusion concerning the nature of the companion. They obtain a
mass of 120$^{+167}_{-43}$\mjup\ for HD~43848~b, with a best fit inclination of 
$12\degr\pm7\degr$ and a best fit ascending node of $288\degr\pm22\degr$. This makes the
companion most likely a low-mass star, in contrast to our analysis which makes it most
likely a high-mass brown dwarf.
\cite{sozzetti10} use the original version of the {\it Hipparcos} Catalogue, not the version
of van Leeuwen (\cite{vanleeuwen07}) as we do. Also, the $\chi^2$ map presented in \cite{sozzetti10} seems
to be in error, since the $\chi^2$ values are different for ascending nodes of 0\degr\
and 360\degr, respectively.

One might at first think that with a period of about 6.5~years, the {\it Hipparcos} data, extending
over little more than three years, would not span a fraction of the orbit which is large enough 
to derive useful constraints on the astrometric orbit. However, due to the high eccentricity
and the resulting fast motion of the components during periastron, a rather large part of the
orbit is covered by the {\it Hipparcos} data; just a small fraction around apastron is not traced.

However, a complication arises from the large uncertainty in the period as determined from
radial velocities, which amounts to 2.3~years and which is a sizable fraction of the period
itself. Following the orbit back in time to the {\it Hipparcos} epoch, the orbital phase is completely 
uncertain. Our analysis does not take the uncertainties of the spectroscopic elements into account.
This is a good approximation for most systems with smaller periods and a good observing record,
but not for HD~43848~b. So if the spectroscopic elements change, so do the inclination, 
ascending node and mass derived here.

In \cite{minniti09} it is stated that HD~43848 has another low-mass stellar companion at a
large projected separation, allegedly discovered by Eggenberger et al.\ (\cite{eggenberger07}). 
However, Eggenberger et al.\ (\cite{eggenberger07})
found a companion with matching mass and projected separation not around HD~43848, but around
HD~43834. Most likely, the companion cited in \cite{minniti09} is not real, but due to a mix
up of HD identifiers.

\subsection{HD~110833 (HIP~62145)}
\label{hd110833}

The brown dwarf candidate HD~110833~b, with a minimum mass of about 17~\mjup\ and a period 
of about 271~days, was discovered by \cite{mayor97} via Doppler spectroscopy. 
The {\it Hipparcos} intermediate astrometric data in their original version were already
analyzed in terms of an orbiting companion by Halbwachs et al.\ (\cite{halbwachs00}), 
who concluded that
the companion was actually a low-mass star with a mass of 0.137$\pm$0.011~\msun.
A similar conclusion was reached by \cite{zucker01}, using the same {\it Hipparcos} data
and spectroscopic parameters as Halbwachs et al.\ (\cite{halbwachs00}); they derive a mass of 
0.134$\pm0.011$~\msun, in perfect agreement with the result of Halbwachs et al.\ (\cite{halbwachs00}).
The best fit inclination in \cite{zucker01} is 7\fdg76; Halbwachs et al.\ (\cite{halbwachs00}) did not
give inclination values.

Interestingly, an astrometric orbital fit for the HD~110833 system was already presented 
in the {\it Hipparcos} Catalogue itself, without the need for input of the spectroscopic parameters.
Assuming a circular orbit (although the orbit is rather eccentric as derived from the
Doppler data) all other orbital paramters were determined astrometrically.
The derived period closely matches the spectroscopically determined period; the derived
inclination is 48\fdg85$\pm$10\fdg20. 

With the new version of the {\it Hipparcos} intermediate astrometric data, we also detect
the astrometric signature of the companion. We obtain a best fit mass of 
102~\mjup; the 3$\sigma$ confidence region for the mass extends from 62 to 141~\mjup,
corresponding to inclinations between 7\fdg7 and 16\fdg8. This is compatible with the
solutions based on the original version of the {\it Hipparcos} Catalogue and confirms that the
companion is most likely a low-mass star.

\subsection{HD~80606 (HIP~45982), 83~Leo~B (HIP~55848) and HD~178911~B (HIP~94075)}
\label{hd80606}

For HD~80606, 83~Leo~B and HD~178911~B the analysis of the {\it Hipparcos}
intermediate astrometric data did not yield useful results. All three stars 
are known binaries, and the contribution of the other component is recognizable
in the data via huge $\chi^2$ values.

HD~80606 is a visual binary star with a separation
of about 30\arcsec. The other component is HD~80607 (HIP~45983), with comparable
spectral type and brightness as the primary. The {\it Hipparcos} data
yielded individual solutions for both components, but they are both marked as 
'duplicity induced variables', as also noted and discussed by \cite{naef01}.
The astrometric solutions carry very large errors in both versions of the 
{\it Hipparcos} Catalogue, so that it is clear that before fitting for a planetary
companion around one of the components the much larger signal of the stellar
component must be subtracted.

83~Leo~B is also a visual binary star; the other component is HIP~55846. The
components are solved as double stars in both versions of the {\it Hipparcos}
Catalogue, but the 
solution is ambiguous according to the notes, and the solutions are flagged 
as being uncertain. The $\chi^2$ value in the original version is rather good,
but 7\% of the data had to be rejected. In contrast, the $\chi^2$ value in the new
version of the {\it Hipparcos} Catalogue is rather large, but only 1\% of the data were 
rejected. 

Similarly, HD~178911~B is also a visual binary star; the other component is HIP~94076. 
A large percentage of the data had to be rejected to obtain a satisfactory solution
in the original version of the {\it Hipparcos} Catalogue. In the new version of the {\it Hipparcos}
Catalogue, the fraction of rejected data is much smaller, at the expense of a
rather large $\chi^2$ value.

\section{Summary and Discussion}
\label{disc}

We have modeled the astrometric orbits of 310 substellar companion candidates
around 258 stars, which had all been previously detected with the radial 
velocity method, using
the {\it Hipparcos} intermediate astrometric data based on the new reduction 
of the {\it Hipparcos} raw data by van Leeuwen (\cite{vanleeuwen07}).
We have obtained the following results:
\\
(1) For all but 67 of the examined companions, we are able to derive an
upper limit for the companion mass, even if the astrometric orbit is not
detected in the data (see Table~\ref{rvobs}).
\\
(2) For nine~companions, the derived $3\sigma$ upper mass limits
are in the planetary mass regime, establishing the planetary nature of these
companions. These planets are, in order of increasing upper mass limits:
$\epsilon$~Eri~b, $\upsilon$~And~d, 47~UMa~b, HD~10647~b,
$\mu$~Ara~b, $\beta$~Gem~b (Pollux~b), HD~147513~b, $\epsilon$~Ret~b, and 
$\upsilon$~And~c. 
\\
(3) Another 75 companions have $3\sigma$ upper mass limits in the brown dwarf
mass range, so that the substellar nature is established (see Fig.~\ref{massfig} 
or Table~\ref{rvobs} for those systems). Two of those (HD~137510~b and
HD~168443~c) have minimum masses also in the brown dwarf mass regime, so that
they are established brown dwarfs.
\\
(4) Even if the astrometric orbit cannot be detected, rather good constraints
on the mass can be derived for a few systems. One example is $\upsilon$~And~d,
for which the lower mass limit based on radial velocities is 4.13~\mjup, and the 
3$\sigma$ upper mass limit based on astrometry is 8.3~\mjup.
\\
(5) For 20 companions, the astrometric orbit could be derived, i.e.\ we get a 
constraint on both the inclination as well as on the ascending node. Those 20
systems are:
HD~18445~b,
BD~$-$04~782~b,
HD~43848~b,
6~Lyn~b,
HD~87883~b,
$\gamma^1$~Leo~A~b,
HD~106252~b,
HD~110833~b,
HD~112758~b,
HD~131664~b,
$\iota$~Dra~b,
$\rho$~CrB~b,
HD~156846~b,
HD~164427~b,
HD~168443~c,
HD~169822~b,
HD~184860~b,
HD~190228~b,
HD~217580~b,
and $\gamma$~Cep~b.
\\
(6) Among the 20~companions for which we could derive astrometric orbits, 
two turn out to be planets (HD~87833~b and $\iota$~Dra~b), eight are brown dwarfs, 
and ten are low-mass stars, judging from the best-fit masses.
\\
(7) The results are in good agreement with astrometric orbits determined by HST/FGS
astrometry for the following companions: $\epsilon$~Eri~b, $\upsilon$~And~c and d,
HD~33636~b, HD~136118~b, 55~Cnc~d, and GJ~876~b.

It is expected that the number of planetary companions detected astrometrically 
will increase dramatically in the future. The PRIMA instrument 
(Quirrenbach et al.\ \cite{quirrenbach98}, Delplancke \cite{delplancke08}) will
enable
differential astrometry with an accuracy of 10--50~microarcseconds, so that it should
be possible to derive precise parameters for many of the systems for which only 
upper mass limits could be derived here. Furthermore, it should enable 
astrometric discoveries of long-period planets around nearby stars which 
are not detectable with current radial velocity precision (see Launhardt et al.\
\cite{launhardt08}). The first 
astrometrically discovered brown dwarf was announced by \cite{pravdo05} and
to our knowledge is so far still the only substellar object discovered 
astrometrically. Future space missions such as SIM Planetquest 
(Unwin et al.\ \cite{unwin08}) and GAIA (Casertano et al.\ \cite{casertano08})
will discover planets and brown dwarfs around nearby stars in large numbers
and will dramatically add to our knowledge about extrasolar planets.

\acknowledgements
We kindly thank Rainer K\"ohler for his help in the early phases of this project,
and Viki Joergens as well as an anonymous referee for useful comments on the manuscript.

% long table number 1
\longtab{1}{
\begin{longtable}{rrclrrrrrr}
\caption{\label{rvobs} Upper mass limits for substellar companion candidates
detected via radial velocities.}\\
\hline\hline\noalign{\smallskip}
  &  & notes &  & period & $m_{2}\sin i  $  & $i_{\mbox{\tiny min}}$ 
& $i_{\mbox{\tiny max}}$ & $m_{\mbox{\tiny 2,max}}$ & $m_{\mbox{\tiny 2,min}}$ \\
\multicolumn{1}{c}{\raisebox{1ex}[-1ex]{designation}} & 
\multicolumn{1}{c}{\raisebox{1ex}[-1ex]{HIP no.}} &
(see text) & \raisebox{1ex}[-1ex]{ref.} & [days]
& [\mjup] & [\degr] & [\degr] & [\mjup] & [\mjup] \\[1.2ex]
\hline\noalign{\smallskip}
\endfirsthead
\caption{continued.}\\
\hline\hline\noalign{\smallskip}
  &  & notes &  & period & $m_{2}\sin i  $  & $i_{\mbox{\tiny min}}$ 
& $i_{\mbox{\tiny max}}$ & $m_{\mbox{\tiny 2,max}}$ & $m_{\mbox{\tiny 2,min}}$ \\
\multicolumn{1}{c}{\raisebox{1ex}[-1ex]{designation}} & 
\multicolumn{1}{c}{\raisebox{1ex}[-1ex]{HIP no.}} &
(see text) & \raisebox{1ex}[-1ex]{ref.} & [days]
& [\mjup] & [\degr] & [\degr] & [\mjup] & [\mjup] \\[1.2ex]
\hline\noalign{\smallskip}
%\hline
\endhead
\hline
\endfoot
          HD 142 b &    522 &    & (1)       &   350.3        &  1.31   &     0.6 &   176.9 &    132.6 &         \\
         HD 1237 b &   1292 &    & (1)       &   133.71       &  3.37   &     4.8 &   175.7 &     46.0 &         \\
         HD 1461 b &   1499 &    & (2)       &     5.7727     &  7.6    &     0.0 &   180.0 &          &         \\
         HD 2039 b &   1931 &    & (1)       &  1120.         &  6.11   &     1.8 &   178.6 &    287.4 &         \\
    BD --17 0063 b &   2247 &    & (3)       &   655.6        &  5.1    &     2.6 &   176.0 &    121.6 &         \\
         HD 2638 b &   2350 &    & (1)       &     3.44420    &  0.477  &     0.0 &   180.0 &          &         \\
         HD 3651 b &   3093 &    & (4)       &    62.218      &  0.229  &     0.2 &   179.7 &     61.2 &         \\
         HD 4113 b &   3391 &    & (5)       &   526.62       &  1.56   &     0.5 &   179.5 &    221.7 &         \\
         HD 4208 b &   3479 &    & (1)       &   828.0        &  0.804  &     0.8 &   178.1 &     58.6 &         \\
         HD 4308 b &   3497 &    & (6)       &    15.560      &  0.0467 &     0.0 &   180.0 &          &         \\
         HD 4203 b &   3502 &    & (1)       &   431.88       &  2.07   &     0.2 &   179.7 &    740.0 &         \\
         HD 5319 b &   4297 &    & (7)       &   674.6        &  1.94   &     0.3 &   179.3 &    438.8 &         \\
         HD 5388 b &   4311 &    & (8)       &   777.         &  1.96   &     1.0 &   179.0 &    124.3 &         \\
         HD 6434 b &   5054 &    & (1)       &    21.9980     &  0.397  &     0.0 &   180.0 &          &         \\
         HD 8574 b &   6643 &    & (1), (4)  &   227.0        &  1.80   &     0.9 &   179.2 &    146.6 &         \\
         HD 9446 b &   7245 &    & (9)       &    30.052      &  0.70   &     0.0 &   180.0 &          &         \\
         HD 9446 c &   7245 &    & (9)       &   192.9        &  1.82   &     0.3 &   179.8 &    651.5 &         \\
  $\upsilon$ And b &   7513 & *  & (1), (10) &     4.617136   &  0.672  &     0.3 &   179.8 &    202.1 &         \\
  $\upsilon$ And c &   7513 & *  & (1), (10) &   241.33       &  1.92   &     8.8 &   172.2 &     14.2 &         \\
  $\upsilon$ And d &   7513 & *  & (1), (10) &  1278.1        &  4.13   &    29.6 &   131.5 &      8.3 &         \\
        HD 10647 b &   7978 &    & (1)       &  1003.         &  0.93   &     7.5 &   174.2 &      9.3 &         \\
        HD 10697 b &   8159 &    & (4)       &  1075.2        &  6.21   &     8.7 &   169.7 &     42.0 &         \\
        HD 11506 b &   8770 &    & (11)      &  1405.         &  4.74   &     2.9 &   177.1 &     98.6 &         \\
        HD 11977 b &   8928 &    & (1)       &   711.0        &  6.5    &     7.0 &   171.1 &     54.7 &         \\
        HD 11964 b &   9094 &    & (1), (10) &  1945.         &  0.622  &     0.7 &   179.0 &     51.1 &         \\
        HD 11964 c &   9094 &    & (10)      &    37.910      &  0.0788 &     0.0 &   180.0 &          &         \\
        HD 12661 b &   9683 &    & (1), (10) &   262.709      &  2.30   &     1.8 &   177.8 &     80.0 &         \\
        HD 12661 c &   9683 &    & (1), (10) &  1708.         &  1.92   &     2.4 &   177.2 &     48.2 &         \\
        HD 13189 b &  10085 &    & (12)      &   471.6        & 14.     &     0.0 &   180.0 &          &         \\
        HD 13445 b &  10138 &    & (1)       &    15.76491    &  3.91   &     3.8 &   177.1 &     81.2 &         \\
         GJ 1046 b &  10812 &    & (13)      &   168.848      & 26.85   &    23.4 &   166.8 &    137.1 &         \\
          79 Cet b &  12048 &    & (1)       &    75.523      &  0.260  &     0.1 &   179.9 &    144.5 &         \\
        30 Ari B b &  12184 &    & (14)      &   335.1        &  9.88   &     3.8 &   174.8 &    162.2 &         \\
        HD 16417 b &  12186 &    & (15)      &    17.24       & 22.1    &     0.0 &   180.0 &          &         \\
        HD 16175 b &  12191 &    & (16)      &   990.         &  4.4    &     2.3 &   177.5 &    115.1 &         \\
          81 Cet b &  12247 &    & (17)      &   952.7        &  5.3    &     1.8 &   178.2 &    176.5 &         \\
        HD 16760 b &  12638 &    & (18)      &   466.47       & 13.13   &     2.7 &   178.5 &    812.9 &         \\
     $\iota$ Hor b &  12653 &    & (1)       &   302.8        &  2.08   &     7.4 &   173.5 &     18.4 &         \\
        HD 17156 b &  13192 &    & (19)      &    21.21663    &  3.22   &     0.1 &   179.9 &   4567.0 &         \\
        HD 18445 b &  13769 &    & (20)      &   554.58       & 44.     &    42.8 &    55.9 &    132.2 &    44.6 \\
                   &        &    &           &                &         &   100.7 &   159.2                      \\
       HIP 14810 b &  14810 &    & (1), (21) &     6.673855   &  3.88   &     0.1 &   179.9 & $>$ 5000 &         \\
       HIP 14810 c &  14810 &    & (1), (21) &   147.730      &  1.28   &     0.1 &   179.8 &   1049.3 &         \\
       HIP 14810 d &  14810 &    & (21)      &   952.         &  0.570  &     0.2 &   179.8 &    161.4 &         \\
        HD 19994 b &  14954 &    & (4)       &   466.2        &  1.37   &     2.9 &   175.2 &     27.2 &         \\
        HD 20782 b &  15527 &    & (1)       &   585.860      &  1.78   &     0.9 &   178.4 &    116.4 &         \\
        HD 20868 b &  15578 &    & (3)       &   380.85       &  1.99   &     0.4 &   179.5 &    335.5 &         \\
  $\epsilon$ Eri b &  16537 & *  & (1), (22) &  2502.1        &  0.86   &     8.8 &   171.4 &      6.1 &         \\
        HD 23127 b &  17054 &    & (23)      &  1214.         &  1.5    &     0.4 &   179.7 &    273.2 &         \\
        HD 23079 b &  17096 &    & (1)       &   730.6        &  2.45   &     4.3 &   175.7 &     33.5 &         \\
        HD 23596 b &  17747 &    & (1), (4)  &  1561.         &  7.71   &     7.7 &   171.8 &     59.2 &         \\
     BD --04 782 b &  19832 &    & (20)      &   716.68       & 47.     &    10.5 &    14.8 &    319.4 &   212.6 \\
  $\epsilon$ Ret b &  19921 &    & (1)       &   428.1        &  1.56   &     6.2 &   174.3 &     13.9 &         \\
        HD 27894 b &  20277 &    & (1)       &    17.9910     &  0.618  &     0.0 &   180.0 &          &         \\
        HD 28185 b &  20723 &    & (1), (4)  &   385.9        &  5.59   &     2.6 &   178.1 &    185.2 &         \\
  $\epsilon$ Tau b &  20889 &    & (24)      &   594.9        &  7.6    &     7.0 &   172.1 &     63.4 &         \\
       HD 283750 b &  21482 & *  & (20)      &     1.787992   & 50.     &     0.1 &   179.9 &   4321.6 &         \\
        HD 29587 b &  21832 &    & (20)      &  1474.9        & 41.     &    25.4 &   152.2 &     97.8 &         \\
        HD 30177 b &  21850 &    & (1)       &  2770.         & 10.45   &     4.0 &   177.1 &    229.0 &         \\
       HD 285968 b &  21932 &    & (25)      &     8.7836     &  0.026  &     0.0 &   180.0 &          &         \\
        HD 30562 b &  22336 &    & (26)      &  1157.         &  1.29   &     2.2 &   178.0 &     38.4 &         \\
        HD 33283 b &  23889 &    & (1)       &    18.1790     &  0.330  &     0.0 &   180.0 &          &         \\
        HD 32518 b &  24003 &    & (27)      &   157.54       &  3.04   &     0.5 &   179.6 &    610.0 &         \\
        HD 33636 b &  24205 & *  & (1)       &  2127.7        &  9.28   &     2.9 &   176.8 &    207.3 &         \\
        HD 33564 b &  25110 &    & (1)       &   388.0        &  9.1    &    26.6 &   162.5 &     30.6 &         \\
        HD 37124 b &  26381 &    & (1)       &   154.46       &  0.64   &     0.2 &   179.8 &    275.2 &         \\
        HD 37124 c &  26381 &    & (1)       &  2295.00       &  0.683  &     0.2 &   179.7 &    230.9 &         \\
        HD 37124 d &  26381 &    & (1)       &   843.60       &  0.624  &     0.7 &   179.3 &     53.9 &         \\
       $\pi$ Men b &  26394 &    & (1)       &  2151.         & 10.27   &    20.3 &   150.6 &     29.9 &         \\
        HD 37605 b &  26664 &    & (1)       &    54.23       &  2.86   &     0.2 &   179.8 &   1915.0 &         \\
        HD 38529 b &  27253 & *  & (1), (10) &    14.31020    &  0.856  &     0.1 &   179.9 &    533.2 &         \\
        HD 38529 c &  27253 & *  & (1), (10) &  2146.1        & 13.1    &     7.7 &   172.6 &    105.3 &         \\
        HD 40307 b &  27887 &    & (28)      &     4.3115     &  0.0132 &     0.0 &   180.0 &          &         \\
        HD 40307 c &  27887 &    & (28)      &     9.620      &  0.0217 &     0.0 &   180.0 &          &         \\
        HD 40307 d &  27887 &    & (28)      &    20.46       &  0.0289 &     0.0 &   180.0 &          &         \\
      HD 41004 A b &  28393 &    & (1)       &   963.         &  2.6    &     2.0 &   177.7 &     80.1 &         \\
      HD 41004 A c &  28393 &    & (1)       &     1.328300   & 18.4    &     0.4 &   179.7 & $>$ 5000 &         \\
        HD 40979 b &  28767 &    & (1), (4)  &   264.15       &  4.01   &     2.8 &   176.2 &     85.3 &         \\
        HD 43848 b &  29804 & *  & (29)      &  2371.         & 25.     &    12.0 &    53.6 &    125.0 &    29.6 \\
                   &        &    &           &                &         &   124.4 &   166.4                      \\
        HD 43691 b &  30057 &    & (30)      &    36.96       &  2.49   &     0.1 &   179.9 &   2675.7 &         \\
        HD 45364 b &  30579 &    & (31)      &   226.93       &  0.1872 &     0.2 &   179.9 &    103.4 &         \\
        HD 45364 c &  30579 &    & (31)      &   342.85       &  0.6579 &     0.2 &   179.4 &    183.5 &         \\
        HD 45350 b &  30860 &    & (1)       &  1003.         &  1.18   &     0.9 &   179.1 &     83.1 &         \\
        HD 45652 b &  30905 &    & (32)      &    43.6        &  0.47   &     0.0 &   180.0 &          &         \\
           6 Lyn b &  31039 &    & (33)      &   874.774      &  2.21   &     1.3 &    13.9 &    104.2 &     9.2 \\
        HD 46375 b &  31246 &    & (1)       &     3.023573   &  0.226  &     0.0 &   180.0 &          &         \\
        HD 47186 b &  31540 &    & (34)      &     4.0845     &  0.0717 &     0.0 &   180.0 &          &         \\
        HD 47186 c &  31540 &    & (34)      &  1353.6        &  0.3506 &     0.6 &   179.7 &     66.1 &         \\
        HD 47536 b &  31688 &    & (1)       &   712.13       &  5.20   &     5.6 &   175.3 &     65.2 &         \\
        HD 48265 b &  31895 &    & (29)      &   762.         &  1.2    &     0.4 &   179.5 &    190.3 &         \\
        HD 49674 b &  32916 &    & (1), (35) &     4.9437     &  0.115  &     0.0 &   180.0 &          &         \\
        HD 50499 b &  32970 &    & (1)       &  2480.         &  1.75   &     1.6 &   178.6 &     76.0 &         \\
        HD 50554 b &  33212 &    & (1)       &  1224.         &  4.46   &     7.8 &   173.1 &     38.0 &         \\
        HD 52265 b &  33719 &    & (1)       &   119.290      &  1.09   &     0.6 &   179.2 &    107.7 &         \\
        HD 60532 b &  36795 &    & (36)      &   201.3        &  1.03   &     1.5 &   178.4 &     41.0 &         \\
        HD 60532 c &  36795 &    & (36)      &   604.         &  2.46   &     6.7 &   173.0 &     21.3 &         \\
        HD 63454 b &  37284 &    & (1)       &     2.817822   &  0.385  &     0.0 &   180.0 &          &         \\
     $\beta$ Gem b &  37826 &    & (37)      &   589.7        &  2.9    &    17.7 &   166.7 &     12.8 &         \\
        HD 65216 b &  38558 &    & (1)       &   613.         &  1.22   &     0.8 &   178.5 &     92.8 &         \\
        HD 66428 b &  39417 &    & (1)       &  1973.         &  2.82   &     1.0 &   179.2 &    216.1 &         \\
        HD 68988 b &  40687 &    & (1)       &     6.27711    &  1.86   &     0.0 &   180.0 &          &         \\
        HD 69830 b &  40693 &    & (38)      &     8.667      &  0.032  &     0.0 &   180.0 &          &         \\
        HD 69830 c &  40693 &    & (38)      &    31.56       &  0.037  &     0.0 &   180.0 &          &         \\
        HD 69830 d &  40693 &    & (38)      &   197.         &  0.0569 &     0.2 &   179.8 &     17.4 &         \\
        HD 70642 b &  40952 &    & (1)       &  2068.         &  1.97   &     3.9 &   176.8 &     35.4 &         \\
        HD 72659 b &  42030 &    & (1), (4)  &  3383.         &  3.15   &     1.4 &   178.2 &    140.4 &         \\
        HD 73267 b &  42202 &    & (3)       &  1260.         &  3.06   &     2.0 &   177.6 &     94.0 &         \\
        HD 73256 b &  42214 &    & (1)       &     2.54858    &  1.87   &     0.0 &   180.0 &          &         \\
        HD 73526 b &  42282 &    & (1)       &   187.499      &  2.04   &     0.2 &   179.8 &   1048.6 &         \\
        HD 73526 c &  42282 &    & (1)       &   376.879      &  2.26   &     0.2 &   179.6 &    822.1 &         \\
        HD 73534 b &  42446 &    & (39)      &  1770.         &  1.103  &     0.2 &   179.8 &    381.7 &         \\
           4 UMa b &  42527 &    & (40)      &   269.3        &  7.1    &     4.4 &   175.2 &     96.2 &         \\
        HD 74156 b &  42723 &    & (41)      &    51.645      &  1.80   &     0.2 &   179.8 &    640.9 &         \\
        HD 74156 b &  42723 &    & (1), (4)  &  2473.         &  8.06   &     3.8 &   175.3 &    128.2 &         \\
        HD 74156 c &  42723 &    & (1), (4)  &   346.6        &  0.40   &     0.1 &   179.8 &    202.5 &         \\
        HD 75289 b &  43177 &    & (1)       &     3.509267   &  0.467  &     0.0 &   180.0 &          &         \\
          55 Cnc b &  43587 & *  & (1), (42) &    14.65162    &  0.824  &     0.2 &   179.8 &    270.1 &         \\
          55 Cnc c &  43587 & *  & (1), (42) &    44.3446     &  0.169  &     0.1 &   179.9 &     92.1 &         \\
          55 Cnc d &  43587 & *  & (1), (42) &  5218.         &  3.835  &     2.8 &   178.5 &    157.2 &         \\
          55 Cnc e &  43587 & *  & (1), (42) &     2.81705    &  0.034  &     0.0 &   180.0 &          &         \\
          55 Cnc f &  43587 & *  & (42)      &   260.00       &  0.144  &     0.2 &   179.8 &     43.4 &         \\
        HD 75898 b &  43674 &    & (7)       &   418.2        &  2.51   &     0.2 &   179.5 &    992.8 &         \\
        HD 76700 b &  43686 &    & (1)       &     3.97097    &  0.233  &     0.0 &   180.0 &          &         \\
        HD 80606 b &  45982 & *  & (4)       &   111.429      &  3.91   &     0.0 &   180.0 &          &         \\
        HD 81040 b &  46076 &    & (1)       &  1001.7        &  6.9    &     5.6 &   173.0 &     72.9 &         \\
        HD 81688 b &  46471 &    & (43)      &   184.02       &  2.7    &     0.5 &   179.5 &    343.7 &         \\
        HD 82943 b &  47007 &    & (1)       &   219.5        &  1.81   &     1.6 &   178.1 &     65.6 &         \\
        HD 82943 c &  47007 &    & (1)       &   439.2        &  1.74   &     3.2 &   177.7 &     44.2 &         \\
        HD 83443 b &  47202 &    & (1), (35) &     2.985625   &  0.400  &     0.0 &   180.0 &          &         \\
        HD 86081 b &  48711 &    & (1)       &     2.13750    &  1.50   &     0.0 &   180.0 &          &         \\
        HD 86226 b &  48739 &    & (44)      &  1534.         &  1.5    &     0.5 &   179.5 &    177.8 &         \\
        HD 86264 b &  48780 &    & (26)      &  1475.         &  7.0    &     1.4 &   178.2 &    307.3 &         \\
    BD --08 2823 b &  49067 &    & (45)      &     5.60       &  0.045  &     0.0 &   180.0 &          &         \\
    BD --08 2823 c &  49067 &    & (45)      &   237.6        &  0.33   &     0.1 &   179.9 &    204.0 &         \\
        HD 87883 b &  49699 &    & (26)      &  2754.         &  1.78   &     4.9 &    34.9 &     21.4 &     3.1 \\
                   &        &    &           &                &         &   154.2 &   173.0                      \\
        HD 88133 b &  49813 &    & (1)       &     3.41587    &  0.299  &     0.0 &   180.0 &          &         \\
        HD 89307 b &  50473 &    & (1), (26) &  2157.         &  1.78   &     1.8 &   177.8 &     57.3 &         \\
$\gamma^1$ Leo A b &  50583 & *  & (46)      &   428.5        &  8.78   &   133.4 &   175.6 &    120.2 &    12.2 \\
        HD 89707 b &  50671 &    & (20)      &   297.708      & 58.     &    17.6 &   148.8 &    207.8 &         \\
        HD 89744 b &  50786 &    & (1), (4)  &   256.78       &  8.44   &     3.8 &   174.1 &    135.1 &         \\
        HD 91669 b &  51789 &    & (47)      &   497.5        & 30.6    &     4.2 &   175.8 &    579.1 &         \\
        HD 92788 b &  52409 &    & (1)       &   325.81       &  3.67   &     2.5 &   178.0 &    113.6 &         \\
        HD 93083 b &  52521 &    & (1)       &   143.58       &  0.368  &     0.2 &   179.8 &    118.6 &         \\
          47 UMa b &  53721 & *  & (1), (4)  &  1076.6        &  2.45   &    16.5 &   164.2 &      9.1 &         \\
          TW Hya b &  53911 &    & (48)      &     3.56       &  1.2    &     0.0 &   180.0 &          &         \\
        HD 96167 b &  54195 &    & (16)      &   498.9        &  0.68   &     0.1 &   179.8 &    418.2 &         \\
        HD 99109 b &  55664 &    & (1)       &   439.3        &  0.502  &     0.2 &   179.9 &    302.7 &         \\
        83 Leo B b &  55848 & *  & (1)       &    17.0431     &  0.109  &     0.0 &   180.0 &          &         \\
       HD 100777 b &  56572 &    & (49)      &   383.7        &  1.16   &     0.3 &   179.8 &    445.1 &         \\
          GJ 436 b &  57087 &    & (1), (35) &     2.643859   &  0.0682 &     0.0 &   180.0 &          &         \\
       HD 101930 b &  57172 &    & (1)       &    70.46       &  0.299  &     0.1 &   179.9 &    176.0 &         \\
       HD 102117 b &  57291 &    & (1), (35) &    20.8079     &  0.172  &     0.0 &   180.0 &          &         \\
       HD 102195 b &  57370 &    & (1), (50) &     4.113775   &  0.45   &     0.0 &   180.0 &          &         \\
       HD 102272 b &  57428 &    & (51)      &   127.58       &  5.9    &     0.1 &   179.9 & $>$ 5000 &         \\
       HD 102272 c &  57428 &    & (51)      &   520.         &  2.6    &     0.1 &   179.9 &   2570.9 &         \\
       HD 104985 b &  58952 &    & (1), (43) &   199.505      &  8.3    &     2.4 &   178.0 &    258.2 &         \\
       HD 106252 b &  59610 &    & (1), (4)  &  1531.         &  6.92   &     7.2 &   173.5 &     63.6 &         \\
       HD 107148 b &  60081 &    & (1)       &    48.056      &  0.210  &     0.0 &   180.0 &          &         \\
       HD 107383 b &  60202 &    & (52)      &   326.03       & 19.4    &     8.7 &   168.8 &    132.3 &         \\
       HD 108147 b &  60644 &    & (1)       &    10.8985     &  0.261  &     0.0 &   180.0 &          &         \\
       HD 108874 b &  61028 &    & (1), (10) &   394.48       &  1.34   &     0.3 &   179.7 &    334.0 &         \\
       HD 108874 c &  61028 &    & (1), (10) &  1680.         &  1.064  &     0.4 &   179.7 &    231.2 &         \\
       HD 109749 b &  61595 &    & (1)       &     5.23947    &  0.277  &     0.0 &   180.0 &          &         \\
       HD 110014 b &  61740 &    & (53)      &   835.477      & 10.9    &     5.5 &   175.7 &    150.9 &         \\
       HD 110833 b &  62145 & *  & (20)      &   271.165      & 17.     &     7.8 &    15.6 &    139.5 &    66.4 \\
       HD 111232 b &  62534 &    & (1)       &  1143.         &  6.84   &    16.7 &   168.9 &     36.3 &         \\
       HD 112758 b &  63366 &    & (20)      &   103.258      & 34.     &     6.6 &    13.9 &    355.6 &   149.8 \\
       HD 114386 b &  64295 &    & (1)       &   938.         &  1.34   &     1.8 &   178.4 &     49.9 &         \\
       HD 114762 b &  64426 &    & (1)       &    83.8881     & 11.68   &     4.1 &   176.7 &    233.0 &         \\
       HD 114783 b &  64457 &    & (1), (4)  &   493.7        &  1.10   &     2.0 &    43.2 &     32.5 &     1.5 \\
                   &        &    &           &                &         &   131.4 &   178.0                      \\
       HD 114729 b &  64459 &    & (1)       &  1114.         &  0.95   &     1.6 &   178.7 &     43.4 &         \\
          61 Vir b &  64924 &    & (54)      &     4.2150     &  0.016  &     0.0 &   180.0 &          &         \\
          61 Vir c &  64924 &    & (54)      &    38.021      &  0.057  &     0.1 &   179.9 &     29.9 &         \\
          61 Vir d &  64924 &    & (54)      &   123.01       &  0.072  &     0.3 &   179.8 &     20.3 &         \\
          70 Vir b &  65721 & *  & (1)       &   116.6884     &  7.49   &    13.2 &   170.3 &     45.5 &         \\
       HD 117207 b &  65808 &    & (1)       &  2597.         &  1.88   &     1.3 &   179.1 &    132.4 &         \\
       HD 117618 b &  66047 &    & (1)       &    25.827      &  0.178  &     0.0 &   180.0 &          &         \\
       HD 118203 b &  66192 &    & (1)       &     6.13350    &  2.14   &     0.0 &   180.0 &          &         \\
      $\tau$ Boo b &  67275 &    & (1)       &     3.312463   &  4.13   &     0.8 &   179.2 &    338.1 &         \\
       HD 121504 b &  68162 &    & (1)       &    63.330      &  1.22   &     0.2 &   179.8 &    478.5 &         \\
       HD 125612 b &  70123 &    & (11)      &   510.         &  3.5    &     1.0 &   178.2 &    240.0 &         \\
       HD 127506 b &  70950 &    & (20)      &  2599.0        & 36.     &    14.2 &   166.2 &    164.9 &         \\
       HD 128311 b &  71395 &    & (1), (4)  &   454.2        &  1.45   &     2.9 &   177.1 &     29.2 &         \\
       HD 128311 c &  71395 &    & (1), (4)  &   923.8        &  3.24   &     7.5 &   171.1 &     25.2 &         \\
       HD 129445 b &  72203 &    & (44)      &  1840.         &  1.6    &     0.2 &   179.8 &    575.1 &         \\
       HD 130322 b &  72339 &    & (1), (4)  &    10.7085     &  1.04   &     0.0 &   180.0 &          &         \\
       HD 132406 b &  73146 &    & (30)      &   974.         &  5.61   &     4.8 &   176.9 &    110.5 &         \\
       HD 131664 b &  73408 &    & (3)       &  1951.         & 18.15   &   153.9 &   171.5 &    131.6 &    42.3 \\
          23 Lib b &  74500 &    & (1), (55) &   258.19       &  1.59   &     0.8 &   178.0 &    118.7 &         \\
          23 Lib c &  74500 &    & (55)      &  5000.         &  0.82   &     0.2 &   179.8 &    365.4 &         \\
          11 UMi b &  74793 &    & (27)      &   516.22       & 11.20   &     6.0 &   174.2 &    113.1 &         \\
       HD 136118 b &  74948 & *  & (1), (4)  &  1187.3        & 11.60   &    13.8 &   172.7 &     95.3 &         \\
          GJ 581 b &  74995 &    & (1), (56) &     5.36874    &  0.049  &     0.0 &   180.0 &          &         \\
          GJ 581 c &  74995 &    & (1), (56) &    12.9292     &  0.0169 &     0.0 &   180.0 &          &         \\
          GJ 581 d &  74995 &    & (1), (56) &    66.80       &  0.0223 &     0.0 &   180.0 &          &         \\
          GJ 581 e &  74995 &    & (56)      &     3.14942    &  0.006  &     0.0 &   180.0 &          &         \\
     $\iota$ Dra b &  75458 &    & (1)       &   511.098      &  8.82   &    29.0 &   133.0 &     18.3 &         \\
       HD 137510 b &  75535 &    & (1)       &   804.9        & 22.7    &    18.4 &   150.9 &     73.9 &         \\
       HD 139357 b &  76311 &    & (57)      &  1125.7        &  9.76   &     4.3 &   173.5 &    143.6 &         \\
       HD 140913 b &  77152 &    & (20)      &   147.956      & 46.     &    11.4 &   167.5 &    257.1 &         \\
       HD 330075 b &  77517 &    & (1)       &     3.387730   &  0.624  &     0.0 &   180.0 &          &         \\
    $\kappa$ CrB b &  77655 &    & (33)      &  1261.94       &  2.01   &     5.1 &   175.1 &     23.8 &         \\
       HD 141937 b &  77740 &    & (1)       &   653.2        &  9.7    &    11.0 &   166.2 &     51.7 &         \\
       HD 142415 b &  78169 &    & (1)       &   386.3        &  1.69   &     0.5 &   179.0 &    231.5 &         \\
      $\rho$ CrB b &  78459 & *  & (1)       &    39.8449     &  1.093  &     0.3 &     0.6 &    219.5 &   108.8 \\
       HD 143361 b &  78521 &    & (29)      &  1086.         &  3.0    &     1.1 &   178.8 &    181.4 &         \\
       HD 142022 b &  79242 &    & (1)       &  1928.         &  4.5    &     4.2 &    49.1 &    102.2 &     4.6 \\
                   &        &    &           &                &         &   102.2 &   177.3                      \\
          14 Her b &  79248 &    & (1)       &  1754.0        &  4.98   &    10.9 &   164.0 &     26.8 &         \\
       HD 145377 b &  79346 &    & (3)       &   103.95       &  5.76   &     0.8 &   179.0 &    542.3 &         \\
         HAT-P-2 b &  80076 &    & (58)      &     5.63341    &  8.64   &     0.0 &   180.0 &          &         \\
       HD 147018 b &  80250 &    & (59)      &    44.236      &  2.12   &     0.2 &   179.6 &    856.0 &         \\
       HD 147018 c &  80250 &    & (59)      &  1008.         &  6.56   &     6.6 &   175.8 &     94.4 &         \\
       HD 147513 b &  80337 &    & (1)       &   528.4        &  1.18   &     5.0 &   175.0 &     13.6 &         \\
       HD 148427 b &  80687 &    & (26)      &   331.5        &  0.96   &     4.8 &   176.6 &    265.0 &         \\
       HD 149026 b &  80838 &    & (1)       &     2.87598    &  0.360  &     0.0 &   180.0 &          &         \\
       HD 150706 b &  80902 &    & (1)       &   264.9        &  0.95   &     0.9 &   178.3 &     66.1 &         \\
       HD 149143 b &  81022 &    & (1)       &     4.07       &  1.33   &     0.0 &   180.0 &          &         \\
       HD 152079 b &  82632 &    & (44)      &  2097.         &  3.0    &     0.3 &   179.8 &   1960.7 &         \\
          GJ 649 b &  83043 &    & (60)      &   598.3        &  0.328  &     0.9 &   179.1 &     21.6 &         \\
       HD 154345 b &  83389 &    & (61)      &  3342.         &  0.947  &     1.7 &   178.1 &     32.6 &         \\
       HD 153950 b &  83547 &    & (3)       &   499.4        &  2.73   &     0.9 &   177.7 &    202.8 &         \\
       HD 155358 b &  83949 &    & (62)      &   195.0        &  0.89   &     0.5 &   179.5 &    122.4 &         \\
       HD 155358 c &  83949 &    & (62)      &   530.3        &  0.504  &     0.5 &   179.4 &     56.8 &         \\
       HD 154672 b &  83983 &    & (63)      &   163.94       &  4.96   &     0.6 &   179.3 &    587.0 &         \\
       HD 154857 b &  84069 &    & (1), (23) &   409.         &  1.8    &     0.3 &   179.4 &    386.3 &         \\
       HD 156668 b &  84607 &    & (64)      &     4.645      &  0.0131 &     0.0 &   180.0 &          &         \\
       HD 156846 b &  84856 &    & (5)       &   359.51       & 10.45   &     2.3 &    17.6 &    475.0 &    29.1 \\
                   &        &    &           &                &         &   157.6 &   178.4                      \\
          GJ 674 b &  85523 &    & (65)      &     4.6938     & 11.09   &     0.0 &   180.0 &          &         \\
       HD 159868 b &  86375 &    & (23)      &   986.         &  1.7    &     0.5 &   179.4 &    193.2 &         \\
       $\mu$ Ara b &  86796 &    & (1), (66) &   643.25       &  1.676  &     7.9 &   170.0 &     12.2 &         \\
       $\mu$ Ara c &  86796 &    & (1), (66) &  4205.8        &  1.814  &     1.9 &   178.0 &     55.0 &         \\
       $\mu$ Ara d &  86796 &    & (1), (66) &     9.6386     &  0.0332 &     0.0 &   180.0 &          &         \\
       $\mu$ Ara e &  86796 &    & (66)      &   310.55       &  0.5219 &     1.5 &   179.1 &     32.4 &         \\
       HD 162020 b &  87330 &    & (1)       &     8.428198   & 15.0    &     0.4 &   179.6 & $>$ 5000 &         \\
       HD 164922 b &  88348 &    & (1)       &  1155.         &  0.360  &     0.9 &   179.3 &     28.2 &         \\
       HD 164604 b &  88414 &    & (44)      &   606.4        &  2.7    &     1.1 &   178.6 &    151.0 &         \\
       HD 164427 b &  88531 &    & (67)      &   108.55       & 46.4    &     7.3 &    33.6 &    444.7 &    85.8 \\
       HD 167042 b &  89047 &    & (33)      &   420.77       &  1.70   &     1.2 &   178.2 &     80.1 &         \\
       HD 167665 b &  89620 &    & (68)      &  4385.         & 50.3    &    22.7 &   140.1 &    137.0 &         \\
       HD 168443 b &  89844 & *  & (1), (10) &    58.11212    &  8.01   &     1.9 &   179.4 &   1232.5 &         \\
       HD 168443 c &  89844 & *  & (1), (10) &  1748.2        & 18.1    &    16.7 &   163.4 &     64.7 &         \\
       HD 168746 b &  90004 &    & (1)       &     6.4040     &  0.248  &     0.0 &   180.0 &          &         \\
          42 Dra b &  90344 &    & (57)      &   479.1        &  3.88   &     3.0 &   177.4 &     88.8 &         \\
       HD 169822 b &  90355 &    & (69)      &   292.1        & 27.2    &   172.9 &   176.2 &    543.5 &   252.5 \\
       HD 169830 b &  90485 &    & (1)       &   225.62       &  2.9    &     2.1 &   178.2 &     99.5 &         \\
       HD 169830 c &  90485 &    & (1)       &  2100.         &  4.1    &     5.2 &   175.1 &     49.3 &         \\
       HD 170469 b &  90593 &    & (11)      &  1145.         &  0.67   &     0.4 &   179.7 &    161.8 &         \\
       HD 171238 b &  91085 &    & (59)      &  1523.         &  2.6    &     0.6 &   179.5 &    373.7 &         \\
       HD 173416 b &  91852 &    & (70)      &   323.6        &  2.7    &     0.7 &   179.6 &    406.2 &         \\
       HD 175541 b &  92895 &    & (71)      &   297.3        &  0.61   &     0.0 &   180.0 &          &         \\
       HD 175167 b &  93281 &    & (44)      &  1290.         &  7.8    &     6.6 &   174.2 &     79.8 &         \\
       HD 177830 b &  93746 &    & (1)       &   410.1        &  1.53   &     0.9 &   179.6 &    225.2 &         \\
     HD 178911 B b &  94075 & *  & (1), (4)  &    71.484      &  7.03   &     0.1 &   179.9 & $>$ 5000 &         \\
       HD 179079 b &  94256 &    & (39)      &    14.476      &  0.0866 &     0.0 &   180.0 &          &         \\
       HD 179949 b &  94645 &    & (1)       &     3.092514   &  0.916  &     0.0 &   180.0 &          &         \\
       HD 181720 b &  95262 &    & (8)       &   956.         &  0.37   &     0.1 &     3.4 &    217.9 &     6.1 \\
                   &        &    &           &                &         &   176.5 &   179.9                      \\
       HD 181433 b &  95467 &    & (34)      &     9.3743     &  0.024  &     0.0 &   180.0 &          &         \\
       HD 181433 c &  95467 &    & (34)      &   962.0        &  0.64   &     0.9 &   179.5 &     76.8 &         \\
       HD 181433 d &  95467 &    & (34)      &  2172.         &  0.54   &     0.4 &   179.8 &    155.4 &         \\
       HD 183263 b &  95740 &    & (1), (10) &   626.5        &  3.67   &     1.9 &   179.2 &    299.3 &         \\
       HD 183263 c &  95740 &    & (10)      &  3070.         &  3.57   &     0.7 &   179.5 &    514.9 &         \\
       HD 231701 b &  96078 &    & (11)      &   141.6        &  1.08   &     0.1 &   179.8 &   1655.7 &         \\
       HD 184860 b &  96471 &    & (69)      &   693.         & 32.0    &   112.0 &   168.8 &    184.3 &    34.6 \\
       HD 185269 b &  96507 &    & (72)      &     6.838      &  0.94   &     0.0 &   180.0 &          &         \\
        16 Cyg B b &  96901 &    & (1)       &   798.5        &  1.68   &     4.0 &   172.4 &     24.3 &         \\
       HD 187123 b &  97336 &    & (1), (10) &     3.0965828  &  0.523  &     0.0 &   180.0 &          &         \\
       HD 187123 c &  97336 &    & (1), (10) &  3810.         &  1.99   &     0.4 &   179.5 &    300.9 &         \\
       HD 187085 b &  97546 &    & (1)       &  1147.0        &  0.98   &     0.7 &   179.1 &     87.7 &         \\
       HD 188015 b &  97769 &    & (1)       &   461.2        &  1.50   &     0.7 &   179.3 &    142.7 &         \\
       $\xi$ Aql b &  97938 &    & (43)      &   136.75       &  2.8    &     1.8 &   179.2 &    214.1 &         \\
       HD 189733 b &  98505 &    & (1)       &     2.21900    &  1.15   &     0.1 &   179.9 &    982.3 &         \\
       HD 190228 b &  98714 &    & (1), (4)  &  1136.1        &  5.93   &     2.5 &    40.8 &    142.9 &     9.1 \\
       HD 190360 b &  98767 &    & (1), (10) &  2915.         &  1.56   &     4.0 &   176.4 &     25.2 &         \\
       HD 190360 c &  98767 &    & (1), (10) &    17.1110     &  0.0600 &     0.0 &   180.0 &          &         \\
       HD 190647 b &  99115 &    & (49)      &  1038.1        &  1.90   &     1.9 &   179.0 &    116.9 &         \\
       HD 190984 b &  99496 &    & (8)       &  4885.         &  3.1    &     0.1 &   179.9 & $>$ 5000 &         \\
       HD 192263 b &  99711 &    & (1)       &    24.3556     &  0.641  &     0.1 &   179.8 &    432.2 &         \\
       HD 192699 b &  99894 &    & (33)      &   345.53       &  2.40   &     0.7 &   179.5 &    301.0 &         \\
       HD 195019 b & 100970 &    & (1), (35) &    18.20163    &  3.70   &     0.2 &   179.7 &   2051.1 &         \\
       HD 196050 b & 101806 &    & (1)       &  1378.         &  2.90   &     3.4 &   177.2 &     61.5 &         \\
       HD 196885 b & 101966 &    & (26)      &  1333.         &  2.58   &     3.4 &   175.2 &     51.9 &         \\
          18 Del b & 103527 &    & (43)      &   993.3        & 10.3    &     7.3 &   171.2 &     82.1 &         \\
     BD +14 4559 b & 104780 &    & (73)      &   268.94       &  1.47   &     0.4 &   179.6 &    276.7 &         \\
       HD 202206 b & 104903 &    & (1)       &   255.870      & 17.3    &     5.5 &   172.4 &    198.3 &         \\
       HD 202206 c & 104903 &    & (1)       &  1383.         &  2.40   &     1.8 &   177.3 &     77.5 &         \\
       HD 204313 b & 106006 &    & (59)      &  1931.         &  4.05   &     2.7 &   175.5 &     92.4 &         \\
          GJ 832 b & 106440 &    & (74)      &  3416.         &  0.64   &     2.8 &   178.0 &     18.5 &         \\
       HD 205739 b & 106824 &    & (63)      &   279.8        &  1.37   &     0.2 &   179.8 &    628.1 &         \\
       HD 208487 b & 108375 &    & (1)       &   130.08       &  0.520  &     0.2 &   179.9 &    256.6 &         \\
       HD 209458 b & 108859 &    & (1)       &     3.52474554 &  0.689  &     0.0 &   180.0 &          &         \\
       HD 210277 b & 109378 &    & (1)       &   442.19       &  1.29   &     2.5 &   177.2 &     30.3 &         \\
          GJ 849 b & 109388 &    & (75)      &  1890.         &  0.82   &     1.8 &   178.2 &     28.1 &         \\
       HD 210702 b & 109577 &    & (33)      &   354.29       &  1.97   &     0.9 &   179.4 &    209.3 &         \\
       HD 212301 b & 110852 &    & (1)       &     2.24572    &  0.396  &     0.0 &   180.0 &          &         \\
       HD 213240 b & 111143 &    & (1)       &   882.7        &  4.72   &     5.9 &   174.9 &     54.1 &         \\
          GJ 876 b & 113020 & *  & (1), (76) &    61.067      &  2.64   &     1.7 &   177.9 &     76.3 &         \\
          GJ 876 c & 113020 & *  & (1), (76) &    30.258      &  0.86   &     0.4 &   179.6 &    109.7 &         \\
          GJ 876 d & 113020 & *  & (1), (76) &     1.93785    &  0.0195 &     0.0 &   180.0 &          &         \\
    $\tau^1$ Gru b & 113044 &    & (1)       &  1311.         &  1.26   &     2.3 &   176.6 &     32.2 &         \\
      $\rho$ Ind b & 113137 &    & (1)       &  1353.         &  2.26   &     5.1 &   175.1 &     26.7 &         \\
       HD 216770 b & 113238 &    & (1)       &   118.45       &  0.65   &     0.2 &   179.8 &    224.9 &         \\
          51 Peg b & 113357 &    & (1)       &     4.230785   &  0.472  &     0.1 &   179.8 &    278.8 &         \\
       HD 217107 b & 113421 &    & (1), (10) &     7.126816   &  1.39   &     0.3 &   179.8 &    556.1 &         \\
       HD 217107 c & 113421 &    & (1), (10) &  4270.         &  2.60   &     1.7 &   178.5 &    104.3 &         \\
       HD 217580 b & 113718 &    & (77)      &   454.66       & 67.     &    32.4 &    67.6 &    131.0 &    72.4 \\
       HD 219828 b & 115100 &    & (50)      &     3.8335     &  0.062  &     0.0 &   180.0 &          &         \\
          14 And b & 116076 &    & (17)      &   185.84       &  4.8    &     1.3 &   179.1 &    332.9 &         \\
       HD 221287 b & 116084 &    & (49)      &   456.1        &  3.09   &     2.8 &   178.3 &    108.3 &         \\
    $\gamma$ Cep b & 116727 &    & (1)       &   905.0        &  1.77   &     3.7 &    15.5 &     28.1 &     6.6 \\
                   &        &    &           &                &         &   171.7 &   174.0                      \\
       HD 222582 b & 116906 &    & (1)       &   572.38       &  7.75   &     8.0 &   175.5 &    105.9 &         \\
       HD 224693 b & 118319 &    & (1)       &    26.730      &  0.71   &     0.0 &   180.0 &          &         \\
\hline\noalign{\smallskip}
\end{longtable}
References:
(1)  \cite{butler06a},
(2)  \cite{rivera10},
(3)  \cite{moutou09},
(4)  \cite{wittenmyer09a},
(5)  \cite{tamuz08},
(6)  \cite{udry06},
(7)  \cite{robinson07},
(8)  \cite{santos10},
(9)  \cite{hebrard10a},
(10) \cite{wright09a},
(11) \cite{fischer07},
(12) \cite{hatzes05},
(13) K\"urster, Endl \& Reffert (\cite{kuerster08}),
(14) \cite{guenther09},
(15) \cite{otoole09},
(16) \cite{peek09},
(17) \cite{sato08b},
(18) \cite{sato09},
(19) \cite{barbieri09},
(20) Halbwachs et al.\ (\cite{halbwachs00}),
(21) \cite{wright09b},
(22) \cite{hatzes00},
(23) \cite{otoole07},
(24) \cite{sato07},
(25) \cite{forveille09},
(26) \cite{fischer09},
(27) \cite{doellinger09b},
(28) \cite{mayor09b},
(29) \cite{minniti09},
(30) \cite{dasilva07},
(31) \cite{correia09},
(32) \cite{santos08},
(33) \cite{bowler10},
(34) \cite{bouchy09},
(35) \cite{wright07},
(36) \cite{desort08},
(37) \cite{reffert06a},
(38) \cite{lovis06}, 
(39) \cite{valenti09},
(40) \cite{doellinger07},
(41) \cite{bean08},
(42) \cite{fischer08},
(43) \cite{sato08a},
(44) \cite{arriagada10},
(45) \cite{hebrard10b},
(46) \cite{han10},
(47) \cite{wittenmyer09b},
(48) \cite{setiawan08}, $T_0\mbox{[JD]} = 2454198.7\pm1.5$ (Setiawan, priv.\ comm., 2008) 
(49) \cite{naef07},
(50) \cite{melo07},
(51) \cite{niedzielski09a},
(52) \cite{liu08},
(53) \cite{demedeiros09},
(54) \cite{vogt10},
(55) \cite{jones10},
(56) \cite{mayor09a},
(57) \cite{doellinger09a},
(58) \cite{loeillet08},
(59) \cite{segransan10},
(60) \cite{johnson10},
(61) \cite{wright08},
(62) \cite{cochran07},
(63) \cite{lopezmorales08},
(64) \cite{howard10},
(65) \cite{bonfils07},
(66) \cite{pepe07},
(67) \cite{tinney01},
(68) \cite{patel07},
(69) \cite{vogt02},
(70) \cite{liu09},
(71) \cite{johnson07},
(72) \cite{johnson06},
(73) \cite{niedzielski09b},
(74) \cite{bailey09},
(75) \cite{butler06b},
(76) \cite{correia10},
(77) \cite{tokovinin94}
} % longtab

\appendix
\section{Apparent Orbit and Astrometric Signature}
\label{apparent}
The astrometric signature of a planetary or brown dwarf companion 
corresponds to the size of the {\it apparent} orbit of the primary star, i.e.\
on the projection of the true orbit into the tangential plane.
Thus, the conversion of the semi-major axes of the true orbit into angular units 
only provides an upper limit for the observable astrometric signature of any given 
companion; the true observable astrometric signature might be considerably smaller,
depending mainly on the eccentricity, the longitude of the periastron and the 
inclination. In the most extreme case, the semi-major axis of the apparent orbit
corresponds to the semi-minor axis of the true orbit only, 
and the semi-minor axis of the apparent orbit could even be identical to zero.

The size of the apparent orbit, i.e.\ its semi-major and semi-minor axis, 
can be calculated from the orbital elements of
the true orbit. Note that most textbooks on double stars cover only the opposite
problem, namely the reconstruction of the true orbit from the observed 
apparent orbit of a visual double star, for which there exist numerous graphical 
and analytical methods (see e.g.\ Heintz \cite{heintz78}).

The Kowalsky method as formulated by Smart (\cite{smart1930}) uses the following general
quadratic equation for a conic section to describe the apparent orbit:
\begin{equation}
\label{conicsection}
Px^2+Qy^2+2Rxy+2Sx+2Ty-1=0 \qquad ,
\end{equation}
where the five parameters $P$, $Q$, $R$, $S$ and $T$ are determined by the five
orbital elements semi-major axis $a$, eccentricity $e$, longitude of periastron $\omega$,
ascending node $\Omega$, and inclination $i$. The equation describes an ellipse
if $P$ and $Q$ have the same sign and are not equal. If $R$ does not equal zero
the ellipse is rotated in the given coordinate system. If $S$ or $T$ do not
equal zero the ellipse is offset from the zero point of the coordinate
system.
According to Smart (\cite{smart1930}), the coefficients in Eq.~\ref{conicsection} are
given by
\begin{eqnarray}
P & = & \frac{1}{a^4 \cos^2i}\left(G^2+\frac{1}{1-e^2}\,B^2\right) \nonumber \\
Q & = & \frac{1}{a^4 \cos^2i}\left(F^2+\frac{1}{1-e^2}\,A^2\right) \nonumber \\
R & = & -\frac{1}{a^4 \cos^2i}\left(F G+\frac{1}{1-e^2}\,A B\right) \\
S & = & \frac{e}{a^2 \cos i}\,G \nonumber \\
T & = & -\frac{e}{a^2 \cos i}\,F \nonumber \qquad ,
\end{eqnarray}
where $A$, $B$, $F$ and $G$ denote the familiar Thiele-Innes constants:
\begin{eqnarray}
A & = & a\, (\cos\omega\cos\Omega - \sin\omega\sin\Omega\cos i) \nonumber \\
B & = & a\, (\cos\omega\sin\Omega + \sin\omega\cos\Omega\cos i) \\
F & = & a\, (-\sin\omega\cos\Omega - \cos\omega\sin\Omega\cos i) \nonumber \\
G & = & a\, (-\sin\omega\sin\Omega + \cos\omega\cos\Omega\cos i) \nonumber  \qquad .
\end{eqnarray}

Note that our scaling factor is different from the one used by
Smart (\cite{smart1930}), because it is convenient for our purposes if the 
absolute term in Eq.~\ref{conicsection} equals $-1$.

In order to determine the semi-major and semi-minor axes of the apparent orbit,
we perform a principal axes transformation. We determine the eigenvalues $\lambda_1$
and $\lambda_2$ of the matrix $\mathcal{M}$
\begin{equation}
\mathcal{M} = \left(\begin{array}{cc} P & R \\ R & Q \end{array}\right) \qquad ,
\end{equation}
which describes the binary quadratic form \ref{conicsection}, as
\begin{eqnarray}
\lambda_1 & = & 1/2\,\left(P+Q-\sqrt{(P-Q)^2+4R^2}\,\right) \\
\lambda_2 & = & 1/2\,\left(P+Q+\sqrt{(P-Q)^2+4R^2}\,\right) \nonumber \qquad .
\end{eqnarray}
Inserting the eigenvalues into the characteristic equation, we obtain
\begin{equation}
\label{characteristic}
\lambda_1 x^2+\lambda_2 y^2 = 1 \qquad ,
\end{equation}
which corresponds to Eq.~\ref{conicsection} in a rotated and shifted coordinate system.
From Eq.~\ref{characteristic} we immediately determine the semi-major axis $a_{\mathrm{app}}$
and the semi-minor axis $b_{\mathrm{app}}$ of the apparent orbit as
\begin{eqnarray}
a_\mathrm{app} & = & 1/\sqrt{\lambda_1} \\
b_\mathrm{app} & = & 1/\sqrt{\lambda_2} \nonumber \qquad .
\end{eqnarray}
The astrometric signature $\alpha$ then directly corresponds to
the semi-major axis of the apparent orbit, $a_{\mathrm{app}}$, converted to angular
units with the help of the parallax $\varpi$:
\begin{equation}
\alpha \mathrm{[mas]} = a_{\mathrm{app}} \mathrm{[AU]} \cdot \varpi \mathrm{[mas]} \qquad .
\end{equation}
Astrometric measurements are sometimes only performed in one dimension at a time;
examples are {\it Hipparcos} or PRIMA (Delplancke \cite{delplancke08}). In the case of PRIMA the
observing direction is given by the baseline orientation, which is flexible to
a certain degree. Thus, it might be advantageous for scheduling purposes to know 
the direction where most of the astrometric signal can be expected, 
i.e.\ the orientation of the apparent orbit. 
The rotation which leads from Eq.~\ref{conicsection} to Eq.~\ref{characteristic}
is characterized by the rotation angle $\varphi$, which can be obtained from
\begin{equation}
\tan 2\varphi = 2 R/(P-Q)
\end{equation}
If the sign of R and P$-$Q is taken into account, this returns an angle $\varphi$ in 
the range between 0 and 360\,\degr\, corresponding to the position angle of the 
semi-major axis of the apparent orbit.

\end{document}